\documentclass[prd,showpacs,nofootinbib,preprintnumbers]{revtex4}

\usepackage{latexsym}
\usepackage{amssymb}
\usepackage{amsmath}
\usepackage{wrapfig}
\usepackage{array}
\usepackage{bm}
\usepackage{subfigure}
\usepackage [english]{babel}
\usepackage[dvips]{epsfig}
\usepackage{slashbox}
\numberwithin{equation}{section}

\newcommand{\ch}{\cosh}
\newcommand{\sh}{\sinh}
\newcommand{\tg}{\tan}
\newcommand{\vth}{\vartheta}
\newcommand{\R}{R_S}

\newcommand{\vk}{\varkappa_D}
\newcommand{\ga}{\gamma}
\newcommand{\al}{\alpha}
\newcommand{\om}{\omega}
\newcommand{\la}{\lambda}
\newcommand{\ka}{\kappa}
\newcommand{\pa}{\partial}
\newcommand{\fr}{\frac}
\newcommand{\s}{\sigma}
\newcommand{\lb}{\label}
\newcommand{\be}{\begin{equation}}
\newcommand{\ee}{\end{equation}}
\newcommand{\ba}{\begin{align}}
\newcommand{\ea}{\end{align}}
\newcommand{\bea}{\begin{eqnarray}}
\newcommand{\eea}{\end{eqnarray}}
\newcommand{\bw}{\begin{widetext}}
\newcommand{\ew}{\end{widetext}}

\newcommand{\e}{{\rm e}}

\newcommand{\nn}{\nonumber}
\newcommand{\zt}{\dot{z}}

\newcommand{\un}{\, ^1 \!}
\newcommand{\y}{{\mathbf{y}}}

\newcommand{\pp}{\ldots}

\newcommand{\inte}{\mathrm{int}}

\newcommand{\nul}{\, ^0 }

\newcommand{\coa}{\xi}

\newcommand{\ths}{}

\newcommand{\od}{\omega}
\newcommand{\ull}{p_{T}}
\newcommand{\uln}{k_{T}}
\newcommand{\comment}[1]{}

\begin{document}

\hfill CCTP-2011-17
\title{Scalar Bremsstrahlung in Gravity-Mediated Ultrarelativistic
Collisions}
%%%%%  Authors  %%%%
\author{Yiannis Constantinou$^\dagger$, Dmitry Gal'tsov$^*$, Pavel Spirin$^{*\dagger}$ and Theodore N. Tomaras$^\dagger$
\thanks{\tt E-mail: galtsov@physics.msu.ru, gkofin@phys.uoa.gr,
salotop@list.ru, tomaras@physics.uoc.gr}} \affiliation{$^*$
Department of Theoretical Physics, Moscow State University, 119899, Moscow, Russian Federation \\
 $^\dagger$ Department of Physics and Institute of Theoretical and
Computational Physics,\\
 University of Crete, 71003, Heraklion, Greece}

\pacs{11.27.+d, 98.80.Cq, 98.80.-k, 95.30.Sf}
\date{\today}

\begin{abstract}
Classical bremsstrahlung of a massless scalar field $\Phi$ is studied in gravity mediated ultra-relativistic
collisions with impact parameter $b$ of two massive point particles in the presence of $d$ non-compact
or toroidal extra dimensions. The spectral and angular distribution of the scalar radiation are analyzed, while
the total emitted $\Phi-$energy is found to be {\it strongly enhanced} by a $d-$dependent power
of the Lorentz factor $\ga$. The direct radiation amplitude from the accelerated particles is shown
to {\it interfere destructively} (in the first two leading ultra-relativistic orders)
with the one due to the $\Phi-\Phi-graviton$
interaction in the frequency regime $\ga/b\lesssim \om \lesssim \ga^2/b$ in all dimensions.
\end{abstract}
\maketitle

\tableofcontents
\section{Introduction and results}\label{intr}
The idea of TeV scale gravity with large extra dimensions (LED) \cite{ablt,ADD,GRW,RS}
has triggered a lot of activity in particle physics and
gravitation theory. One of the most interesting predictions is the
possibility of black hole production in colliders \cite{BH}.
According to Thorne's hoop conjecture (generalized to higher
dimensions \cite{Ida}),  for energies of colliding particles higher
than the $D$-dimensional Planck mass $M_*$ (transplanckian regime)
such black holes should be produced classically for impact
parameters $b\lesssim \R $, where $\R$ is the Schwarzschild radius
associated with the center-of-mass collision energy. To establish
the creation of a black hole in the collision of ultrarelativistic
particles one has to find closed trapped surfaces in the
corresponding space-time. To  this aim, an idea due to Penrose
\cite{Penrose} was put into the form of an elaborated model in $D=4$
by D'Eath and Payne \cite{DEPA} and with extra dimensions by Eardley
and Giddings \cite{Eardley} and further refined in \cite{Yoshino}
(see recent reviews \cite{recent} and references therein). In that
model one is after a solution of Einstein's equations with a
special metric ansatz generalizing the Aichelburg-Sexl metric
\cite{AS} (see also \cite{DH}). The ansatz amounts to replacing the
gravitational field of two ultrarelativistic particles before the
collision  by colliding shock waves, while the collision region is
described by some linear differential equation for a metric
function, which is amenable to construct exact or approximate
solutions. The closed trapped surface emerging in such a solution
for appropriate initial energies and impact parameters leads to an estimate of the produced
black hole mass and its difference from the initial energy is interpreted as the amount
of gravitational radiation produced (see e.g. the recent paper
\cite{Herdeiro:2011ck} and references therein).

The colliding waves model (CWM) of black hole formation is certainly a very nice and perhaps the
simplest possible one, designed to answer an intriguing question about the nature of transplanckian
collisions. It gives the gravitational radiation loss for head-on and almost head-on collisions and
also demonstrates that the black
hole is indeed present in the collision region. However, we would like to
discuss some subtle points concerning the applicability of this model to high energy particle
scattering. In fact, various effects which were not taken into account in the simple
version of CWM such as an extended nature of the colliding particles \cite{GiRy} and their parton
structure \cite{MR}, have already been discussed. It was shown by Meade and Randall \cite{MR} that
taking the finite size into account  leads  to a substantial decrease of the cross-section
predicted by the CWM. It was also emphasized \cite{MR} that an even more
critical effect is the radiative energy loss of the colliding
partons before their energy is trapped inside a black hole horizon.
The CWM seems to be able to clarify these issues, but in a closer
look it probably cannot. In fact, the Aichelburg-Sexl solution is
the limiting form of the linearized gravitational field of ultrarelativistic particle moving with a {\em constant}
velocity. Therefore, presenting colliding particles as plane waves implicitly
assumes that their radiation is negligibly small, otherwise the
particle trajectories should be modified substantially by radiation
friction (for a discussion of radiation reaction in extra dimensions
see \cite{React}). Furthermore, the energy mismatch between the mass
of the black hole and the initial energy of the colliding particles,
interpreted in CWM as radiation loss, is found to be of the order of
the initial energy \cite{Eardley}, which is by no means small.

Therefore, alternative methods of computing radiation losses in
transplanckian collisions seem to be desirable. One such
approach is the one by Amati, Ciafaloni and Veneziano \cite{Venez}
relevant in four dimensions and based on the combination of
string and quantum field theory techniques. Other recent work on this
subject includes \cite{other} using various analytical classical and
semi-classical approaches. In the framework of purely classical
$D=4$ general relativity, numerical simulations were also performed
of collision of two scalar field balls interacting gravitationally
via exact Einstein equations \cite{CP}. Gravitational radiation was
extracted imposing appropriate boundary conditions.  Gravitational
radiation in collisions of higher-dimensional black holes (with
non-compact extra dimensions) was studied numerically in
\cite{Witek:2010xi}.

For the ultrarelativistic scattering in models
with large extra dimensions a crucial question is whether radiation
is enhanced due to the extended phase space associated with extra
dimensions. It was argued by Mironov and Morozov \cite{Miro} that
in the case of synchrotron radiation the expected enhancement can be
damped by beaming of radiation in the forward direction, suppressing
the number of excited Kaluza-Klein modes. In the case of
bremsstrahlung the situation is different, namely in \cite{GKST-2}
it was found, that the  energy loss of ultrarelativistic particles
under non-gravitational scattering at small angle contains an
additional factor $\ga^d$ due to the emission of light massive KK
modes. Qualitatively, this can be explained as follows. For
non-gravitational scattering in flat space the impact parameter $b$,
the radiation frequency $\om$ and the angle of emission
$\vartheta\ll 1$  with respect to the momentum of the fast particle
in the rest frame of the other are related by
 \begin{equation}
 \om b(\vartheta^2+\ga^{-2} )  \lesssim 1.
 \end{equation}  Frequencies near the cut-off frequency
$\om\sim\ga^2/b$ are emitted in the narrow $(D-1)$ dimensional cone
$\vth\lesssim \ga^{-1}$, while intermediate frequencies are emitted
into a wider cone. The main contribution comes from the first
region, and in this case the emitted momenta in directions
transverse to the brane are of the order $\om/\ga$. The number of
such light modes is of the order $(R\om/\ga)^d\sim (R\ga/b)^d$ ($R$
being the size of the compact extra dimensions), giving an extra
factor $\ga^d$ to the radiated power. Analogous enhancement was
reported for gravitational bremsstrahlung in transplanckian collisions \cite{GKST-PLB}, compatible with
the numerical study of \cite{Witek:2010xi}.

The purpose of this paper is to study scalar radiation in gravity-mediated collisions in the presence of
extra dimensions. Our approach was initiated in \cite{GKST-PLB, GKST-EL} where we have briefly
formulated purely classical perturbation theory for  transplanckian
scattering  in ADD valid for small scattering angles, or large
impact parameters. Indeed, as emphasized long ago by many
authors, most notably by 't Hooft \cite{'tHooft:1987rb},
the gravitational force not only becomes dominant at transplanckian
energies, but gravity itself becomes classical, at least in some
significant impact parameter regime.
Thus, the applicability of classical theory in the bremsstrahlung problem at transplanckian energies
seems to have solid theoretical grounds.

More specifically, our approach amounts to solving the two-body
field-mediated problem iteratively. In electrodynamics this is a well known method, allowing to
calculate spectral-angular distributions of bremsstrahlung in the
classical range of frequencies small with respect to the particle
energy $\hbar \om\ll E$. In general relativity this approach was
suggested in \cite{FMA} under the name of ``fast-motion approximation scheme''
and was further developed and called ``post-linear formalism''. It was
applied to the gravitational bremsstrahlung most notably by Kovacs
and Thorne \cite{KT}. We will use a momentum space version of this
approach, developed in \cite{GaGr}, which has the advantage that it allows a fully analytical treatment
of the problem. In \cite{GKST-PLB, GKST-2} we extended this technique to models with
extra dimensions either infinite or compact. In \cite{GKST-2} we
calculated scalar bremsstrahlung radiation in the collision of two ultrarelativistic
point-like particles interacting via a scalar field  in flat
space-time. Here we consider the situation in which the particles interact gravitationally but
emit scalar radiation. This is an intermediate step towards the full treatment of the
gravitational bremsstrahlung presented briefly in \cite{GKST-PLB}. The latter case has additional complications
due to the tensor nature of the radiation field and it will be presented in full detail in a future publication.
Here, we will assume that only one of the colliding particles is coupled to the scalar field,
so that their interaction is purely gravitational. On the other
hand, we will compute only the scalar radiation emitted by
the system. The main novel feature is this case is that the system becomes non-linear due to
scalar-scalar-graviton vertex. As a result, the effective  source of
radiation field in the flat space picture becomes non-local due to contribution of the field stresses.

In four dimensions, it was shown long ago \cite{ggm}, that the contribution from
the high-frequency regime $\ga/b\lesssim \om\lesssim \ga^2/b$ is {\em suppressed by a factor $\gamma^{-4}$ due
to destructive interference } between the local and the non-local
amplitudes, the latter being due to the gravitational interaction of
the mediating field. The remaining radiation is beamed inside the cone with angle $1/\gamma$, it
has characteristic frequencies
$\om \sim {\mathcal O}(\ga/b)$ and emitted energy of order $E\sim \gamma^3$.
\comment{
This suppression is, however, compensated by
$\ga$-proportionality of the gravitational interaction force, so
the total loss has the same $\ga$-dependence as in the case of
non-gravitational scattering \cite{ggm} (up to numerical
coefficient)
 \be\lb{ezprim}
E'_0\sim\frac{(\varkappa_4 m f_4)^2}{b^3} \ga^3.
 \end{equation}
 }
There is in addition a sub-leading component of emitted radiation with frequencies $\om\sim {\mathcal O}(\ga^2/b)$,
which is also beamed and has $E\sim \ga^2$.

In the higher dimensional case the situation is more complicated.  The
destructive interference is also present, but with growing $d$ the
relative contribution of $\om\sim \ga^2/b$ increases faster, than
that of $\om\sim\ga/b$ due to competition of the angular integrals.

The powers of $\gamma$ of the emitted radiation energy in all frequency and angular regimes in the presence of
$d$ extra dimensions are summarized in the Table below.

\vspace{0.5cm}
\begin{center}
\begin{tabular}{|c|c|c|c|c|}
  \hline \backslashbox{$\vartheta$}{$\od$} & $\od\ll \gamma/b$  & $ \od \sim \gamma/b$ &
  $\od \sim \gamma^2/b $& $\od \gg \gamma^2/b$
  \\\hline
  $\hspace{3pt} \gamma^{-1} \hspace{3pt}$ & $\begin{array}{c} \text{\small negligible}\\ \text{\small
   (phase space)} \end{array}$ &
  $ \hspace{1cm} E_d \sim \gamma^3 \hspace{1cm}
 $  &
   $\begin{array}{l} \hspace{0.3cm} E_d \sim \gamma^{d+2}  \hspace{0.2cm} \\
   %z'\text{\small - type insig. (exponential fall-off)}
   \end{array}$
   & $\begin{array}{c}  \text{\small  negligible radiation}\\ \end{array}$
  \\[20pt]   \hline
  1 & $\begin{array}{c} \text{\small negligible}\\ \text{\small  (phase space)} \end{array} $ &
   $ \hspace{0.5cm} E_d \sim \gamma^{d+1} \hspace{0.5cm}  $&
   $\begin{array}{c}  \text{\small negligible radiation}\\  \end{array}$&
    $\begin{array}{c}  \text{\small  negligible radiation}\\  \end{array} $\\[20pt] \hline
\end{tabular}
\end{center}

\vspace{0.5cm}

In the special cases of $d=1$ or 2 extra dimensions there is an extra $\ln \gamma$ in the
expression for the energy emitted
in the regime ($\om\sim \gamma/b$, $\vartheta\sim 1/\gamma$).
The energy is measured in units of $\vk^4 {m'}^2 f^2/b^{3d+3}$, with $m'$ the mass of the
target particle, $\vk$ the $D-$dimensional gravitational coupling, $f$ the scalar coupling of $m$ and $b$ the
impact parameter of the collision and the overall numerical coefficients can be found in the text.

\section{The setup}\label{setup}

\subsection{The action}
The goal here is to calculate within the ADD scenario classical
spin-zero bremsstrahlung in ultra-relativistic gravity-mediated scattering of two
massive point particles $m$ and $m'$. The space-time is assumed to be $M_4\times T^d$, the product of
four-dimensional Minkowski space
and a $d$-dimensional torus, with
coordinates $x^M=(x^\mu,y^i),\, M=0,1,\ldots, D-1, \mu
=0,\dots ,3,\, i= 1,\dots  d$.

Particles move in $M_4$ (the brane) and interact via the
gravitational field $g_{MN}$, which propagates in the whole
space-time $M_4\times T^d$. We also assume the existence of a
massless bulk scalar field $\Phi(x^P)$, which interacts with $m$,
but not with $m'$. The action of the model is symbolically of the
form \begin{equation}\label{Actot} S\equiv S_g+S_\Phi+S_m+S_{m'}\,, \nonumber
\end{equation}and explicitly, in an obvious correspondence, \begin{equation}S\!=\!\!\int\!
d^D x
 \sqrt{|g|} \! \left[\! -\frac{R}{\vk^2}+\frac{1}{2}\,g^{MN}\partial_M\Phi\,
\partial_N\Phi\right] -\frac{1}{2}\!\int\! \!\left[e \, g_{MN} \zt^M
\zt^N \!+\! \frac{(m\!+\!f \Phi)^2}{e} \right]\! d\tau
-\frac{1}{2}\!\int\! \left[e' g_{MN} \zt'^M \zt'^N \!+\!
\frac{m'^2}{e'} \right] \!d\tau'
\end{equation}
with $16\pi G_D\equiv \vk^2$
relating $\vk$ to Newton's constant. Here the ein-beins $e(\tau)$
and $e'(\tau')$ are introduced, which lead to a somewhat unusual
form of interaction with $\Phi$, but it reduces to the standard
non-derivative interaction once they are integrated out. The
constant $f$ is the scalar charge of $m$. To solve the corresponding
coupled equations of motion we shall use perturbation theory with respect to
the scalar and gravitational couplings. To zeroth
order, both gravitational and scalar fields are absent and the two
particles move along straight lines as determined by the initial
conditions. In first order, one takes into account the
(non-radiative Coulomb-like) gravitational and scalar fields
produced by the particles in their zeroth order trajectories. Next,
one computes the first order correction to their trajectories, due
to their first order gravitational interaction (the scalar mutual
interaction vanishes in our set-up). The leading contribution to the
scalar radiation field, of interest here, emitted by the accelerated
particle $m$ is then obtained in the second order of perturbation
theory. This approach allows us to compute consistently the lowest
order radiation in the case of ultrarelativistic collision, when
deviations from straight trajectories is small and the iterative
solution of the coupled particle-field equations of motion is
convergent. The resulting expression  for the radiative energy loss
will be therefore correct only in the leading ultrarelativistic
order.

To carry out such a computation it is sufficient to
restrict oneself to linearized gravity. One writes $
g_{MN}=\eta_{MN}+\vk h_{MN} $ and replaces the Hilbert-Einstein
action by its quadratic part
 \begin{equation}
 S_g=\int\left[ -\frac{1}{4} h^{MN} \square_D h_{MN}
+\frac{1}{4} h \square_D h -\frac12 h^{MN}\partial_M
\partial_N h  +\frac12 h^{MN}\partial_M
\partial_P h^P_N  \right]d^Dx,
\label{FP}
 \end{equation}
 where the Minkowski metric is $\eta_{MN}={\rm diag}
(1,-1,-1,...)$,
 $\square_D\equiv \eta^{MN}\partial_M\partial_N$, raising/lowering
the indices is performed with $\eta_{MN}$ and $h\equiv h^M_M$. To
avoid classical renormalization (which in principle can be treated
along the lines of \cite{React}) we take into account only {\em
mutual } gravitational interaction. At the linearized level the
total gravitational field $h_{MN}$ is the superposition of the
fields $h^m_{MN},\;h^{m'}_{MN}$ due to the particles $m$ and $m'$,
respectively, and of the gravitational field generated by the bulk
scalar $\Phi$. Assuming that the scalar interaction is of the same
order as the gravitational one, the latter contribution to $h_{MN}$
is of higher order and will be neglected here. Therefore, to this
order of approximation
 \begin{equation}
 h_{MN}=h^m_{MN}+h^{m'}_{MN}\,.
 \end{equation}
For simplicity the superscripts $m,\,m'$ will be omitted in what
follows. Thus, $h^m_{MN}$ and $h^{m'}_{MN}$ will be denoted as
$h_{MN}$ and $h'_{MN}$, respectively. Ignoring the self-interaction
and the associated radiation reaction problem, we will consider each
of the particles $m'$ and $m$  as moving in the other's  metric
\begin{equation}
g_{MN}= \eta_{MN} + \varkappa_D h_{MN} \;\; {\rm and} \;\; g'_{MN}= \eta_{MN} +
\varkappa_D h'_{MN}\,,
\label{gg'}
\end{equation}
respectively. Correspondingly, the action for the particle $m$,
which interacts with both gravity and $\Phi$, takes the form
 \begin{equation}
 \label{Acm}
S_m =-\frac{1}{2}\int \left[e (\eta_{MN}+\vk h'_{MN}) \zt^M \zt^N + \frac{(m+f\Phi)^2}{e} \right] d\tau.
 \end{equation}

Similarly, the action for $m'$ is
\begin{equation}
\label{Acm'}
S_{m'} =-\frac{1}{2}\int \left[e' (\eta_{MN}+\vk h_{MN}) \zt'^M \zt'^N +
\frac{m'^2}{e'} \right] d\tau.
\end{equation}

Finally, the action for the scalar field, which propagates in the
gravitational field $h'_{MN}$ of the uncharged particle, expanded to
linearized order of the gravitational field, is
 \begin{equation}
S_{\Phi}=\frac{1}{2}\int \partial^{M}\Phi\,
 \partial^{N}\Phi\,\left[\left(1+ \frac{\varkappa_D}{2}
 h'\right)\eta_{MN} -\varkappa_D h'_{MN}\right]
 d^{D}x\,.
 \end{equation}
 In principle, $\Phi$ propagates in the full gravitational field $h_{MN} + h'_{MN}$ of both particles,
 but  the  singular product of $h_{MN}$ and $\Phi$ generated by the same point
particle $m$ corresponds again to the  self-action problem, which is ignored here. The products
of $h'_{MN}$ due to $m'$ and $\Phi$ due to $m$ does not lead to singularities and correctly
describe the situation.
 %%%%%%%%%%%%%%%%%%%%%%%%%%%%%%%%%%%

\subsection{Equations of motion}
Varying the action
with respect to $z(\tau)$ and $z'(\tau)$
one obtains the linearized geodesic equations of each mass moving in the gravitational field
of the other:
 \begin{equation}\label{9}
\frac{d}{d \tau}\left(e g'_{MN} \zt^N\right)=\frac{e}{2}
g'_{LR,M}\zt^L \zt^R, \qquad\frac{d}{d \tau}\left(e' g_{MN}
\zt'^N\right)=\frac{e'}{2} g_{LR,M}\zt'^L \zt'^R.
 \end{equation}
Variation with respect to the einbeins gives
 \begin{equation}\label{10}
e^{-2}=\frac{g'_{MN} \zt^M \zt^N}{(m+f \Phi)^2}  \qquad
e'^{-2}=\frac{g_{MN} \zt'^M \zt'^N}{m'^2}
 \end{equation}
Substituting this back into the particles' actions
(\ref{Acm}-\ref{Acm'}) one is led to their more familiar form
$$
S_m =-\int  (m+f \Phi) (g'_{MN} \zt^M \zt^N)^{1/2} d\tau, \qquad
S_{m'} =-  m'  \int (g_{MN} \zt'^M \zt'^N)^{1/2} d\tau,
$$
%\end{equation}
from which the scalar field equation is obtained
 \begin{equation}
 \label{sc_f_eq_tot}
\square_D\Phi=-\frac{\vk}{2} h' \square_D\Phi
+\vk h'_{MN}\Phi^{, MN}+f \int (g'_{MN} \zt^M \zt^N)^{1/2}
\delta^D(x-z(\tau)) \, d\tau,
 \end{equation}
Note, once again, that only the
gravitational field due to the uncharged particle $m'$ enters this
equation.

Finally, the linearized Einstein equations for the metric deviation due to the
two particles in the De Donder gauge
$$\pa_N h^{MN}=\frac{1}{2}\, \pa^M h$$
are obtained from (\ref{Acm}, \ref{Acm'}):
 \begin{equation}
 \label{Einm}
  \square_D h^{MN}=-\vk \left( T^{MN}-\eta^{MN}\;\frac{T}{D-2}\right),\quad
  T^{MN}=\int e{\dot z}^M {\dot z}^N
  \frac{\delta^D(x-z(\tau))}{\sqrt{-g'}} d\tau,
 \end{equation}  where $T=T_M^M$ and linearization of the metric factor  is
understood. Similarly,
 \begin{equation}
 \label{Einm'}
  \square_D h'^{MN}=-\vk \left( T'^{MN}-\eta^{MN}\;\frac{T'}{D-2}\right),\quad
  T'^{MN}= \int e' {\dot z}'^M {\dot z}'^N
  \frac{\delta^D(x-z'(\tau))}{\sqrt{-g}} d\tau.
 \end{equation}

To ensure that the particles move on the brane, it is enough to choose
the initial conditions  $y^i(0)=0,\,{\dot y}^i(0)=0$ and similarly
for $m'$. Then, using the equations of motion, it is easy to check
that the entire world-lines will be $x^\mu=z^\mu(\tau),\;
x'^\mu=z'^\mu(\tau)$ and the energy-momentum tensors will only have
brane components $T^{\mu\nu}, T'^{\mu\nu}$. However, the
metric deviations $h^{MN}$ and $h'^{MN}$ will have in addition
diagonal bulk components due to the trace terms in (\ref{Einm}) and (\ref{Einm'}).
%%%%%%%%%%%%%%%%%%%%%%%%%%%%%%%%%%%%%%%%%%%%%%%%
%%%%%%%%%%%%%%%%%%%%%%%%%
\subsection{Iterative solution}

Even in linearized gravity the relativistic two-body problem
can not be solved exactly, so one has to use some approximation scheme.
With the particle masses $m,\,m'$ taken of the same order and eventually equal,
the model is characterized by three classical length
parameters. Namely, the classical radius of the scalar charge
\cite{GKST-2}:
 \begin{equation}
   r_f=\left(\frac{f^2}{m}\right)^{\frac1{d+1}},
 \end{equation}
the $D$-dimensional gravitational radius of the mass $m$
at rest
 \begin{equation}
   r_g=\left(\vk^2 m\right)^{\frac1{d+1}},
 \end{equation}
 and the Schwarzschild
radius of the black hole, associated with the collision energy
$\sqrt{s}$ \cite{GKST-PLB}:
 \begin{equation}
   r_S=\frac{1}{\sqrt{\pi}} \left[ \frac{8\Gamma
\left( \frac{d+3}{2}\right)}{d+2}\right]^{
\frac{1}{d+1}}\left(\frac{G_D
 \sqrt{s}}{c^4}\right)^{\frac1{d+1}}.
 \end{equation}
  In the (initial) rest frame of the mass $m'$ one has $\sqrt{s}=2mm'\ga$,
where $\ga=1/\sqrt{1-v^2}$ is the Lorentz factor of the collision,
$v$ being the relative velocity of the colliding particles. So
 \begin{equation}
   r_S\sim r_g\ga^\nu,\quad \nu=\frac1{2(d+1)}.
 \end{equation}
It will be assumed that the  parameters $r_g$ and $r_f$ are of the same order, and both
much smaller than the impact parameter $b$:
 \begin{equation}
   r_g\sim r_f \ll b\, \ga^{-2\nu},
 \end{equation}
  or, in terms of $r_S$ \cite{GKST-PLB}:
 \begin{equation}
 b\gg r_S \ga^{\nu}.
 \end{equation}
Under this condition, as will be shown below, the deviation of the metric from unity in the rest frame
of $m'$ is small, i.e. $\vk h_{MN}\zt'^M \zt'^N \ll 1$, which is necessary for the validity of the present
perturbative treatment.

\subsubsection{The formal expansion and zeroth order equations}

The next step is to solve these equations iteratively. For the particle-$m$ world-line one writes
 \begin{equation}
 \label{snickers}
z^{M} = \nul z^{M}+\,\un z^{M} + \pp,\qquad
 \nul z^{M}=u^{M}\tau  + b^M,
 \end{equation}
where the order is denoted by a left superscript and to zeroth order the particle moves in Minkowski
space-time with constant velocity $u^M$, and with $ \nul z^{M}(0)=b^M$ another constant vector.
Both vectors $u^M,\, b^M$ will be assumed to lie on the brane, i.e. to have
only $\mu$-components and, in addition, to be orthogonal $b^\mu u_\mu$=0.
It will be shown that as a consequence of the equations of motion $\un z^{M}$ also lies on the brane.
However, it is convenient to keep $D$-dimensional notation in all intermediate steps.

For the particle-$m'$ one writes similarly
 \begin{equation}\label{snickers1}
z'^{M} = \nul z'^{M}+ \un z'^{M} + \pp,\qquad
 \nul z'^{M}=u'^{M}\tau,
 \end{equation}
assuming that at $\tau=0$ the particle is at the origin.
We choose to work in the rest frame of $m'$, and specify the coordinate axes
on the brane so that $ u'^\mu= (1,0,0,0),\; u^\mu=
\gamma(1,0,0,v),\;
 \gamma=1/\sqrt{1-v^2}$, and $b^\mu=(0,b,0,0)$, where $b$ is the impact parameter. When
needed, one may think of the brane-localized vectors as
$D$-dimensional vectors with zero bulk components, e.g. $u^M=(u^\mu, 0,\ldots,0)$.

In a similar fashion, the bulk scalar is expanded formally as:
 \begin{equation}
 \Phi=\nul\Phi+\un\Phi+\pp.
 \end{equation}
Substitute in (\ref{sc_f_eq_tot}) and set $\vk=0$ to obtain for the leading contribution
to $\Phi$ the equation
\begin{equation}
\label{scag}
\square_D  \nul \Phi=f\int \delta^D(x-u\tau- b) \, d\tau.
\end{equation}

The ein-beins are also expanded
\begin{align}
\label{eexpand}
 e = \nul e+ \un e + \pp, \qquad e' = \nul e'+ \un e' + \pp.
 \end{align}
According to (\ref{10}) one obtains in zeroth order
 \begin{equation}
\nul e=m+f\; \nul\Phi, \qquad \nul
 e'=m'.
 \end{equation}

Finally, for the metrics one writes
 \begin{equation}  h_{MN}=\nul h_{MN}+\un h_{MN}+\pp,
 \end{equation}
and similarly for $h'_{MN}$.
The leading order contributions to the metrics are then obtained from Eqs. (\ref{Einm}, \ref{Einm'})
with the zeroth order source on the right hand side, i.e. with
\begin{equation}\label{Tzero}
\nul T^{MN}=m\int  \delta^D(x-\nul z(\tau))\, u^{M} u^{N} d\tau,\quad
\nul T'^{MN}=
m\int  \delta^D(x-\nul z'(\tau)) \, u'^{M} u'^{N} d\tau.
\end{equation}

To calculate the leading order scalar bremsstrahlung it will be sufficient to know only the
zeroth order term $\nul h_{MN}$ of $h_{MN}$. So, in the sequel only $\nul h_{MN}$ will appear
and to simplify the notation, its left superscript will be omitted.

\subsubsection{The first order equations}

To derive the equations for the first order
corrections to the particle world-lines one has to collect first order
terms in the expansions of the embedding functions $z^M,\, z'^M$ and
the einbeins  $e, \,e'$  and choose suitable gauge condition to fix the $\tau, \, \tau'$ reparametrization
symmetries. From Eqs. (\ref{10}) one finds for the first order
corrections of the einbeins:
 \begin{equation}
\un e= - e_0 \left(\vk h'_{MN} u^M u^N + 2 \un \zt_M  u^M\right
),\qquad \un e'=- e'_0 \left(\vk h_{MN} u'^M u'^N + 2 \un \zt'_M
u'^M\right ).
 \end{equation}
The reparametrization freedom allows us to fix $\un e$ and  $\un
e'$ arbitrarily. We first substitute these expansions into
Eq.(\ref{9}), collect all the first order terms and then choose the gauge fixing conditions
$\un e=\un e'=0$, that is
 \begin{equation}
 \label{Gagm}
 \vk h'_{MN} u^M u^N + 2\un \zt_M  u^M=0
 \end{equation}  in the equation for $m$, and
 \begin{equation}
 \label{Gagm'}
 \vk h_{MN} u'^M u'^N + 2\un \zt'_M  u'^M=0
 \end{equation}  in the equation for $m'$. The resulting
equations for the first corrections to the particle trajectories read
\begin{align}
\label{z1m}
\Pi^{MN} \un \ddot{z}_{N}&=-\varkappa_D  \Pi^{MN}\left(h'_{NL,R}-
 \frac{1}{2} h'_{LR,N}\right)u^{L}u^{R},\\
 \Pi'^{MN} \un \ddot{z'}_{N}&=-\varkappa_D  \Pi'^{MN}\left(h_{NL,R}-
 \frac{1}{2}h_{LR,N}\right)u'^{L}u'^{R},
\end{align}
where the projectors onto the space transverse to the world-lines
are
 \begin{equation}    \Pi^{MN}= \eta^{MN}-u^{M}u^{N},\quad \Pi'^{MN}=
  \eta^{MN}-u'^{M}u'^{N}\,,
 \end{equation}
whose presence  demonstrates explicitly that only the transverse
perturbations of the world-lines are physical.

 %%%%%%%%%%%%%%%%%%%%%%%%%%%%%%%%%%%%%%%%%%

\subsubsection{The second order equation for $\Phi-$radiation}

The radiative component of the bulk scalar arises in the next order of
iterations and is given by $\un\Phi$. With appropriate combination of terms on the right hand side,
Eqn. (\ref{sc_f_eq_tot}) is written as:
 \begin{equation}
 \label{Phij}
 \square_D\un\Phi(x,y)=j(x,y)\equiv\rho(x,y)+\sigma(x,y),
 \end{equation}
where the first term is localized on the world-line of the radiating particle $m$
\begin{equation}\label{rholoc}
 \rho(x,y)=-f\int \un z^{\mu}(\tau)\, \partial_\mu \delta^4(x-u\tau-b)\,\delta^d(\y) d\tau,
 \end{equation}
while the second is the non-local current
\begin{equation}
\sigma (x,y) = \varkappa_D \;\partial_M
\left(h'^{MN}\,\partial_N \nul \Phi -
\frac{1}{2} h' \, \partial^{M} \, \nul\Phi \right),
\end{equation}
with both $h'^{MN}$ and $\nul \Phi$ having support in the bulk. This
current arises from non-linear terms in (\ref{sc_f_eq_tot}) due to the
interaction of the bulk scalar with gravity. It has to be emphasized
that it is non-zero outside the world-line not only on the brane but
also in the bulk. Note that the decomposition into the local and
non-local parts is ambiguous in the sense that part of the non-local
term can be cast into a local form using the field equations. But the
total source $j$ never reduces to a local form as a whole.

%%%%%%%%%%%%%%%%%%%%%%%%%%%%%%%%%%%%%%%%%%%%%

\subsection{The solution for  $\nul\Phi$, $h_{MN}$ and $\un z^M$ in $M_4\times T^d$}

Our notation and conventions for Kaluza-Klein decomposition and
Fourier transformation as applied to the ADD scenario with the
transverse directions being circles with radii equal to $R$, are given
in Appendix \ref{KKnotation}. It is important to stress at this point that in
classical perturbation theory the interaction is described as an
exchange of {\em interaction modes} (the classical analogs of
virtual gravitons), but contrary to the Born amplitudes, where the
simple pole diagrams diverge at high transverse momenta \cite{GRW},
here the sum over these modes contains an intrinsic cut-off.
Therefore the classical elastic scattering amplitude is finite and,
furthermore, it   reproduces the result of the eikonal method, if
the eikonal is computed in the stationary phase approximation
\cite{GKST-EL}. In other words, classical calculations in ADD
correspond to non-perturbative ones in quantum theory (the eikonal
method is equivalent to  summation of the ladder diagrams). As will
be explicitly demonstrated in the present paper, the same is true
for bremsstrahlung. Specifically, it will be shown that the
effective number of interaction modes is finite due to the cut-off
and of order $(R/b)^d$. Also, the summation over the {\em emission
modes} is cut-off to a finite effective number of order
$(R\ga/b)^d$, which leads to a large extra enhancement factor
$\ga^d$ in ultrarelativistic collisions.

%\subsubsection{The solution for $\nul\Phi$, $h_{MN}$ and $\un z^M$ in Fourier space}

Straightforward Fourier transform of (\ref{scag}) gives
\begin{equation}
\label{phi(1)} \nul \Phi^n(p)= -\frac{2\pi f  \delta(p
u)}{p^2-p_T^2}.
\end{equation}

Similarly, for the $h_{MN}$ and ${h'}_{MN}$, it is enough to Fourier transform
the source terms (\ref{Tzero}) and plug into (\ref{Einm}) and (\ref{Einm'}). One then obtains
\begin{align}
\label{grsola} h^n_{MN}(p)= \frac{2\pi \vk \, m\,
\delta(pu)}{p^2-p_T^2}\e^{i(pb)}\left( u_{M}u_{N} -
\frac{1}{D-2}\eta_{MN} \right),
\end{align}
where $p^2=p_\mu p^\mu,\; pu=p_\mu u^\mu$ and $p_T^i=n^i/R$ is the
quantized momentum vector in the transverse directions. To get
${h'}^n_{MN}(p)$, one has to replace $m\to m',\, u^M\to u'^M, b\to
0$. Note that these fields do not describe radiation. They simply
represent the scalar and gravitational potentials of the uniformly
moving particles. Formally, this follows from the presence of the
delta factors $\delta(pu)$ with $pu=\gamma(p^0-p_z v)$, from which
it follows that $p_\mu p^\mu=p_z^2(v^2-1)<0$, while the mass-shell
condition for the emitted wave is $p_\mu p^\mu=p_T^2\geqslant 0$.

%%%%%%%%%%%%%%%%%%%%%%%%%%%%%%%%%%%%%%%%%%%%%%%%%%%
Substitution of (\ref{grsola}) into (\ref{z1m}) and integration of the resulting equation gives
\begin{align}
\label{grdef} \un z^{M}(\tau)=-\frac{ i m' \varkappa_D^2}{(2 \pi)^3
V} \sum_{l} \int d^4 p \frac{\delta(pu')}{(p^2-p_T^2)
(pu)}\e^{-i(pb)} \left(\e^{-i(p u) \tau} -1 \right)\left(\gamma
u'^{M}-\frac{1}{d+2} u^{M}-\frac{\gamma_*^2}{2(p u)}
 p^{M}\right)\,,
\end{align}
with $\gamma_*^2\equiv \gamma^2-{(d+2)}^{-1}$ and the
$D$-dimensional vector $p^M=(p^\mu, p_T^i=l^i/R)$. It is easy to
check that the gauge condition (\ref{Gagm}) is satisfied. To ensure
this {\em exactly} one has to keep the small second term in
$\gamma_*^2$, which, however, will be dropped in what follows in
view of  our interest in $\ga\gg 1$. We have chosen the initial
value $\un z^{M}(0)=0$ in order to preserve the meaning of $b^\mu$
as the impact parameter, namely $b^{\mu}=z^\mu(0)-z'^\mu(0)$. Note
that the initial value of $\un \zt^{M}(0)$ is non-zero and is
computed from the gauge condition (\ref{Gagm}).

From (\ref{grdef}) one can prove that the gravitational interaction
does not expel the particles from the brane. Indeed, only the last
$p^M$-term in the last parenthesis has non-zero components
$p_T^i=l^i/R$ orthogonal to the brane. But the remaining expression
is even under the reflection $l^i\to -l^i$ and the sum vanishes giving
$\un z^{i}=0$ \footnote{This is on the average true also quantum
mechanically. The $\Phi$-quanta emission is symmetric on the average
under reflection from the brane and the brane stays on the average
at rest. However, to guarantee transverse momentum conservation in
single $\Phi$ emission in the bulk one should introduce brane
position collective coordinates and deal also with the brane back
reaction.}.

The corresponding solution for the mass $m'$ can be obtained by
interchanging $u^M$ and $u'^M$, replacing $m$ by $m'$ and omitting
$\e^{-i(pb)}$.

%%%%%%%%%%%%%%%%%%%%%%%%%%%%%%%%%%%%%%%%%%%%%%

\subsection{$\Phi-$radiation - Basic formulae}

Finally, $\Phi-$radiation is described by the wave equation (\ref{Phij}), which
in terms of Kaluza-Klein modes is
\begin{equation}
\label{Phin} (\square+k_T^2)\un\Phi^n(x)=j^n(x)\equiv
\rho^n(x)+\sigma^n(x)\,,
\end{equation}
where $k^i_T=n^i/R$, while $\rho^n(x)$ and $\sigma^n(x)$ are
\begin{equation}
\lb{ron}
\rho^n(x)= -f\int \un z^{\mu}(\tau)\, \partial_\mu
\delta^4(x-u\tau-b)d\tau
 \end{equation}
 and
 \begin{equation}
 \lb{son}
 \sigma^n(x)= \frac{\vk}{V}\sum_{l} \partial_\mu  \left({h'}_{l}^{\mu\nu}(x)\,
 \partial_\nu \nul \Phi_{n-l}(x) -
  \frac{1}{2} h'_{l}(x) \, \partial^\mu \, \nul\Phi_{n-l}(x) \right)\,,
\end{equation}
respectively, with
$\un z^\mu$, $\nul\Phi$ and $h'_{\mu\nu}$ given in (\ref{grdef}), (\ref{phi(1)}) and (\ref{grsola}).
The rest of this paper is devoted to the solution of (\ref{Phin}) and the analysis of the spectral
and angular distribution of the emitted $\Phi-$radiation.

%%%%%%%%%%%%%%%%%%%%%%%%%%%%%%%%%%%%%%%%%%%%%%%%%%%

Once the solution of these equations is available, one can compute the energy and momentum radiated
away using the standard formulae of radiation theory.
To compute the momentum loss due to scalar bremsstrahlung emitted
during the collision one considers the world tube with topology $R^{1,3}\times T^d$, with boundary
$\pa\Omega=\Sigma_{\infty}\cup\Sigma_{-\infty}\cup B$ consisting
of two space-like hypersurfaces $\Sigma_{\pm\infty}$ in $R^{1,3}$
at $t=\pm\infty$ and the time-like hypersurface $B$ at
$r\to\infty$ and integrate the difference of the fluxes through
$\Sigma_{\pm\infty}$ to obtain
 \begin{equation}
 \label{flux1}
P^\mu=\int\limits_V d^dy
\Biggl(\,\,\int\limits_{\Sigma_{\infty}}-\int\limits_{\Sigma_{-\infty}}\!\Biggr)
T^{\nu\mu}d^3\Sigma_\nu.
\end{equation}
Here one makes use of the brane components of the energy-momentum
tensor of the bulk scalar (it is easy to show that there is no
radiation flux into the compact dimensions \cite{GKST-2}). Start with
 \begin{equation}
 T^{MN}=\pa^M \Phi \pa^N \Phi -\frac12 \eta^{MN} (\pa\Phi)^2,
  \end{equation} where only $\un\Phi$ has to be taken into account, since $\nul\Phi$ is not
related to radiation. The integral over $B$ is zero due to fall-off
conditions, so the difference of the surface integrals
(\ref{flux1}) can be transformed by Gauss' theorem to the volume
integral \footnote{It was taken into account that $T^d$ has
no boundary}
\begin{equation}\label{flux2}
P^\mu=\int\limits_V d^dy
 \int\limits_{\Omega} \pa_N T^{N\mu }d^4x
 = \int\limits_V d^dy
 \int\limits_{\Omega} \left(\pa^\mu\Phi \right)\;\square_D \Phi \;d^4x.
  \end{equation}
Using (\ref{Phij}) and the retarded Green's function of the
$D-$dimensional D'Alembert operator to solve the wave equation
\begin{equation}
\label{KK1a22_mi1}
 G_{D}(x-x',y-y') =\frac1{(2\pi)^4 V }
 \int d^4 k \, \e^{-ik(x-x')} \sum_n
 \frac{\e^{i k_T(y-y')}}{k^2-k_T^2+i
\epsilon k^0}\,,
\end{equation}
one obtains
\begin{equation}
P^\mu=\frac1{16\pi^3 V}\sum_n \int \frac{d^3 k}{k^0} \,k^\mu \, |j^n(k)|^2\,
 \Big|_{k^0=\sqrt{{\bf k}^2+k_T^2}}\,,
\end{equation}
where $j^n(k)$ is the Fourier-transform of $j^n(x)$ (for precise
definition see Appendix \ref{KKnotation}) and will be referred to as the {\it
radiation amplitude}. With the parametrization $\bf{ k}= |\bf{
k}|(\sin\theta\cos\varphi,\,\sin\theta\sin\varphi,\,\cos\theta)$ one
obtains  for the spectral-angular distribution of the emitted energy
$E=P^0$:
 \begin{equation}
 \label{radiat}
 \frac{dE}{d|{\bf k}| d\Omega_2}=\frac1{16\pi^3 V}\sum_n
{ \bf k}^2|j^n(k)|^2,\quad d\Omega_2=\sin\theta \,d\theta
\,d\varphi.
 \end{equation}
Therefore, the leading order radiation loss is determined by the
Fourier-transform $j^n(k)$ of the source in the four-dimensional
wave equation (\ref{Phij}) for $\un\Phi^n$ on the mass shell of
emitted waves
\begin{equation}
\lb{ms} k_\mu  k^\mu=k_T^2.
\end{equation}
If the impact parameter is small compared to the compactification
radius ($b\ll R$), the summation over KK masses can be replaced by
integration according to (\ref{sumint}), with integration measure
$d^d k_T =k_T^{d-1}\, dk_T \,d\Omega_{d-1}$. Since the radiation
amplitude $j^n$ depends only on $ |k_T|$, integration over the
angles is trivial and gives the volume $\Omega_{d-1}$ of the unit
$d-1-$dimensional sphere. Therefore,
\begin{equation}
\label{radiat1} \frac{dE}{d |\mathbf{k}| d\Omega_2}=
\frac{\Omega_{d-1}}{2(2\pi)^{D-1}} \int\limits_0^\infty {\bf
k}^2k_T^{d-1}\;|j^n(k)|^2\;dk_T.
 \end{equation}
with $n$ in $j^n$ expressed in terms of $k_T$, and $k^0\equiv
\om=\sqrt{{\bf k}^2+ k_T^2}$.

\begin{figure}
\begin{center}
\includegraphics[angle=0,width=15cm]{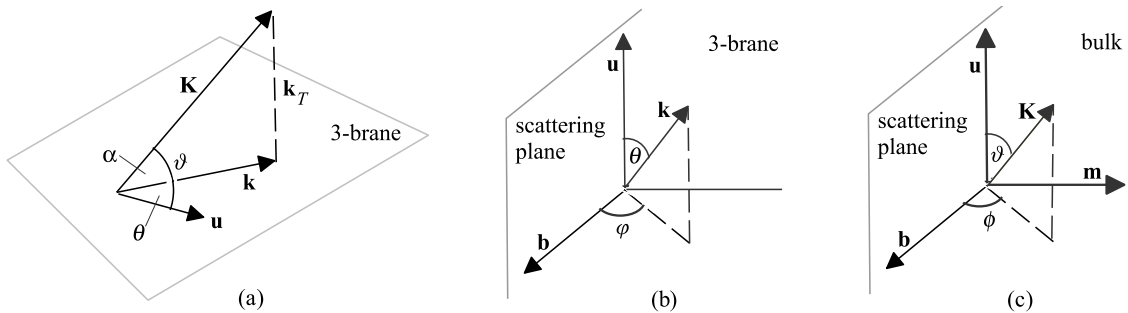}
\caption{The angles in lab frame used in the text.}
\label{branepic}
\end{center}
\end{figure}

When the summation over KK emission modes is replaced by
integration, one can compute the emitted energy using directly
higher dimensional Minkowski coordinates. Take the coordinate system
with angles on the $(D-2)-$dimensional sphere $\Omega_{D-2}$, with
$\vartheta$ the angle between the $(D-1)-$dimensional vectors
$\mathbf{K}\!=\!({\bf k}, k_T^i)$ and $\mathbf{u}$, $\phi$ the
polar angle in the plane perpendicular to ${\bf u}$ varying from 0
(direction of $\mathbf{b}$) to $2 \pi$ (see Figure \ref{branepic})
and integrate over the angles to obtain \cite{GKST-2}
\begin{align}
\label{pafinalmink} \frac{d E}{d\omega  d\Omega_{d+2}}= \frac{
\omega^{d+2} }{2(2 \pi)^{d+3}} |j(k)|^2.
\end{align}
where $\od \equiv(ku')=k^0=\sqrt{{\bf k}^2+ k_T^2}$ is the higher-dimensional frequency.
%%%%%%%%%%%%%%%%%%%%%%%%%%%%%%%%%%%%%%%%%%%%%%%%%

\section{The radiation amplitude}

As we have seen, the radiation amplitude $j^n(k)$ is the sum of a
local $\rho^n(k)$ and a non-local part $\s^n(k)$. These two parts have different
dependence on the frequency and the angle $\theta$ of the emitted
wave with respect to the direction of collision. Both have an
intrinsic cut-off at some (angle-dependent) frequency, which in the
ultrarelativistic case is high compared to the inverse impact
parameter $1/b$. Typically, the two amplitudes cancel each other in
some range of angles and frequencies. To obtain the correct expression for their sum one has to
carefully take into account non-leading contributions.
%%%%%%%%%%%%%%%%%%%%%%%%%%%%%%%
\subsection{The local amplitude}
\label{local}

Fourier-transformation of (\ref{ron})  gives
\begin{align}
\label{loc_varn}
\rho^n(k)=i f e^{i(kb)}\int\limits_{-\infty}^{\infty} \e^{i  (ku) \tau} (k
\un z)\, d\tau
\end{align}
where the scalar products are denoted as $(ab)\equiv a_\mu b^\mu$ and for simplicity the
parentheses will also be omitted when it is not ambiguous. Substitution
of $\un z^{\mu}$ from (\ref{grdef}) and integration over $\tau$ gives \footnote{Note that the unity inside
the first parenthesis of (\ref{grdef}) corresponds to a constant $\un z$ and does not contribute to
radiation. Formally, its contribution to $\rho^n$ vanishes by symmetric integration \cite{GKST-2}.}:
\begin{align}
\label{Clinton01}
\rho^n(k)=  \frac{\varkappa_D^2 \, m' f \e^{i(kb)}}{4 \pi^2 V (k
u)^2}\sum_{l} \left[ k  u \left(\gamma k   u'-\frac{k u}{d+2}\right)
I_{l} -\frac{\gamma_*^2}{2} k_{\mu} I_{l}^{\mu} \right],
\end{align}
where the integrals $I_l$ and $I_l^{\mu}$ are defined by
\begin{align}
I_{l} = \int \frac{ \delta(pu')\, \delta(ku-pu)\,\e^{-i(pb)}}{p^2-\ull^2}\;d^4 p,\quad
I^{\mu}_{l} = \int \frac{ \delta(pu')\, \delta(ku-pu)\, \e^{- i(pb)}}{p^2-\ull^2}
\;p^{\mu} \;  d^4 p,
\end{align}
The sum over $l$ represents the sum over
the interaction modes labeled by the set of integers $l^i$, while the dependence of the
amplitude $\rho^n(k)$ on the vector index $n$, which labels the
emission modes $n^i$, is hidden inside the $k^0-$component of the wave
vector: $k^0=\sqrt{\mathbf{k}^2+\uln^2}$. The integrals are given in \cite{GKST-2} and lead to Macdonald
functions:
\begin{align}
I_{l}  =-\frac{2\pi}{\gamma v}
{K}_{0}(z_l),\quad
I^{\mu}_{l}
 =-\frac{2\pi }{\gamma v b^{2}}\left(b z {K}_{0}(z_{l})\frac{\gamma
u'^{\mu}-u^{\mu}}{\gamma v}+ i \hat {K}_{1}(z_{l})\, b^{\mu}\right),
\end{align}
with
\begin{equation} \lb{z1}
 z\equiv \fr{(ku) b}{\gamma v}, \quad z'\equiv \fr{(ku')b}{\gamma v}  ,\quad
 z_{l}\equiv (z^2+\ull^2 b^2)^{1/2}\,,
 \end{equation}and the hatted Macdonald functions defined by $\hat{K}_{\nu}(x)\equiv x^{\nu}{K}_{\nu}(x)$
and having for $\nu\neq 0$ a finite non-zero limit as $x\to 0$.
Thus, in terms of Macdonald functions the local amplitude is:
\begin{equation}
\label{rhon(k)} \rho^n(k)= -\frac{\varkappa_D^2 m'f }{4 \pi v V
}\, \e^{i(kb)} \sum_{l}
\left\{\left[\left(2-\frac{\ga_*^2}{v^2\ga^2}\right)\frac{
z'}{z}-\frac{2}{\ga}\left(\frac{1}{d+2}-
\frac{\ga_*^2}{2v^2\ga^2}\right)\right]K_0(z_l)
-i\frac{\ga_*^2}{\ga^2}\frac{(kb)}{v^2 \ga z^2} \hat
K_1(z_l)\right\}\,.
\end{equation}

\subsubsection{The $\ga\to\infty$ limit. Mode and frequency cut-offs.}

In the ultrarelativistic limit $\ga\to\infty$ the leading terms of $\rho^n$ are
\begin{align}
\label{Clinton02mm}
\rho^n(k)\simeq -\frac{\varkappa_D^2 m'f }{4 \pi v V }\, \e^{i(kb)}
\sum_{l} \left[\frac{z'}{z}
{K}_{0}(z_{l})- i\frac{(k b)}{\ga z^2}\,\hat{K}_1(z_{l}) -\frac{d+1}{(d+2)\ga^2}\left(\frac{z'}{z}-\frac{\ga d}{d+1}\right) K_0(z_l)
+``\mathcal{O}(\ga^{-2})"\right]
\end{align}
This is a systematic ultra-relativistic expansion in powers of $1/\ga$, modulo the coefficients of the various
Macdonalds as well as the overall factor in front, which
depend on the velocity $v$. However, as will become evident below, this form is
adequate for the following discussion and the computation of the emitted energy to leading order.

Two important general remarks are in order here:

(a) {\it The effective number $N_{\inte}$ of interaction modes.}
The exponential fall-off of the Macdonald functions at large
values of the argument $z_l$ leads to an effective cut-off $N_{\rm
int}$ of the number of interaction modes $l$ in the sum. One can
estimate the radius $\ull^\inte$ of the sphere in the space of
$l^i$, beyond which the modes can be neglected, by setting \begin{equation}(\ull^{\inte})^2\, b^2\sim 1\,, \label{lmax} \end{equation}from which \begin{equation}N_{\rm int}\sim \left(\frac{R}{b}\right)^d\sim
\frac{V}{b^d}. \end{equation}For $N_{\rm int} \gg 1$, which is the case of
interest here, one may use (\ref{sumint}) to obtain \cite{GKST-2}
($Z>0$)
\begin{align}
\label{sum2int}
\frac{1}{V} \sum_{l }\hat{K}_{
\lambda}\left(\sqrt{Z^2+p_T^2 b^2}\right) \simeq
\frac{1}{(2\pi )^{d/2} b^d} \hat{K}_{\lambda+d/2}(Z)
\end{align}
and end up with
\begin{align}
\label{rhon_final} \rho^n(k)\simeq -\frac{\lambda \e^{i(kb)}}{v}
\left\{ \frac{z'}{z} \,\hat K_{d/2}(z) - i\frac{(k b)}{\ga
z^2}\hat{K}_{d/2+1}(z) +\frac{1}{(d+2)\ga} \! \left(d-
\frac{(d+1)z'}{\ga z} \right)  \hat{K}_{d/2}(z)
+``\mathcal{O}(\ga^{-2})"\!\right\},
\end{align}
where
\begin{equation}
\label{lambda}
\lambda\equiv \frac{\vk^2 m' f}{2 (2\pi)^{d/2+1} b^d}\,.
\end{equation}

(b) {\it Angular and frequency characteristics.} The local radiation amplitude above in the
$b\ll R$ limit is expressed solely in terms of Macdonald functions with argument $z$. Later, it will be
shown that the non-local amplitude contains also Macdonald functions but with argument $z'$. The exponential fall-off
of these functions implies
the effective cut-offs $z\sim 1$ and $z'\sim 1$ in the corresponding radiation amplitudes. These, in turn, translate into
angular and frequency characteristics of the corresponding radiation.

Specifically, with $\theta$, $\alpha$ and $\vartheta$ as shown in
Fig. \ref{branepic}, define \begin{equation}\psi\equiv
1-v\cos\theta\cos\alpha=1-v\cos\vartheta \,,
\end{equation}which satisfies
\begin{equation}\frac{z}{z'}=\ga \psi\,, \end{equation}and in the ultrarelativistic
limit varies in the interval $1/2\ga^2 \simeq 1-v \leqslant \psi
\leqslant 1+v \simeq 2$.

Consider the neighborhood of $z\sim 1$ which gives the dominant contribution of the local radiation
amplitude. One has to distinguish various domains of the emission angles.
For small emission angles $\theta, \al$ one has
\begin{equation}
\psi\sim\fr12(\ga^{-2}+\theta^2+\al^2)\,,
\end{equation}
so that (i) inside the {\em small} cone $\theta^2+\al^2\lesssim 1/\ga^2$ one obtains $\psi\sim 1/\ga^2$,
so that the characteristic frequencies
$\om\sim \ga^2/b$. This angle-frequency regime will be called in the sequel {\it the
z-region}. (ii) For $\psi\sim 1$, i.e. for $\alpha, \theta \sim \mathcal{O}(1)$, one obtains $z'\sim z/\ga$,
which implies a low frequency regime $\om\sim 1/b$, whose
contribution to the emitted energy is negligible in view of the relative
smallness of the phase-space factor in (\ref{radiat1}).

To summarize, the above analysis of $\rho^n$ combined with the phase
space factors in (\ref{radiat1}), leads to the conclusion that the
leading contribution of the radiation due to the local amplitude is
{\it beamed}, i.e. directed inside the small cone
$\theta^2+\al^2\lesssim 1/\ga^2$ and has high frequencies $\om \sim
\ga^2/b$.  Radiation with these characteristics will occasionally be
called {\it z-type}.

%%%%%%%%%%%%%%%%%%%

\subsection{The non-local amplitude}

The non-local amplitude obtained from (\ref{son}) by Fourier transform is
\begin{align}
\label{Clinton05} \s^n(k) =\frac{\varkappa_D^2  m' f  (k u')^2}{(2
\pi)^{2}}\,\e^{i (k  b)} J^n(k)\,, \quad J^n(k)\equiv \frac{1}{V}
\sum_{l }J^{nl}(k)\,,
\end{align}
with
\begin{align}
J^{nl}(k)= \int  \,d^4 p \frac{\delta(pu')\delta(ku-pu) e^{-i(pb)}}{
(p^2-p_T^2)\, [ (k-p)^2 - (k_T-p_T)^2] } \,.
 \end{align}
$k^0=\sqrt{\mathbf{k}^2+k_T^2}$ and $\mathbf{k}$ is a
$3-$dimensional vector lying on the $3-$brane,  where $k_T^i=n^i/R$
and $p_T^i=l^i/R$ with integers $\{n^i\}, \{l^i\}$ are
$d-$dimensional discrete vectors corresponding to the emission and
interaction modes, respectively.

Using Feynman parametrization $J^{nl}$ takes the form:
\begin{align}
J^{nl}=\int\limits_0^1  dx \, e^{- i (kb)x} \int d^4 p \frac{
\delta[(pu')+ (ku') x]
\delta[(pu)-(1-x)(ku)]\,\e^{-i(pb)}}{[p^2-(k_T x-p_T)^2]^2}. \nn
\end{align}
Integrating over $p^0$ and splitting
$\mathbf{p}$ into the longitudinal $p_{||}$ and transversal
$\mathbf{p_{\bot}}$ parts, integrate over $p_{||}$. Then,
introducing in $\mathbf{p_{\bot}}$ the spherical coordinates
$d^{2}\mathbf{p_{\bot}}=|\mathbf{p}_\perp|d\Omega_{1}
d|\mathbf{p}_{\bot}|$ and integrating first over the angles and
then over $|\mathbf{p}_{\bot}|$, one obtains
\begin{align}
\label{hhh6} & J^{nl} =\frac{\pi b^2}{\gamma v} \int\limits_0^1 dx\,e^{-i
(kb)x}\, \hat{K}_{-1}(\zeta_{nl})\,,
\end{align}
with
\begin{align}
\label{hhh5} \zeta_{nl}^2(x)= {z'}^2 x^2 + 2\ga z z' x(1-x) + z^2
(1-x)^2 +b^2(k_Tx-p_T)^2 \,.
\end{align}
Again, the summation over $l$ is performed trivially for $b\ll R$
using (\ref{sum2int}) \footnote{Using (\ref{sumint}) the summation
is converted to integration over $d^dl$. One then shifts the
integration variable $\ull'=\ull-x\uln$
and applies (\ref{sum2int}). In the present case $Z$ is not
constant but depends on $x$. However, as it can also be checked
numerically, (\ref{sum2int}) and the subsequent treatment is a
good approximation for any $0\leqslant x
\leqslant 1$, because $Z(x) \gtrsim 1$ in all relevant frequency regimes.}. The result is
\begin{align}
\label{cucu1}
J^{n}(k) =\Lambda_d \int\limits_0^1 dx\,e^{-i(kb)x}\, \hat{K}_{d/2-1}(\zeta_n)  ; \qquad
\Lambda_d\equiv \frac{\pi b^{2-d}}{(2\pi)^{d/2} \gamma v}\,.
\end{align}
with
\begin{align}
\label{cucu2}
\zeta_n^2(x)= {z'}^2 x^2 + 2\ga z z' x(1-x) + z^2 (1-x)^2  \,; \qquad  \zeta_{n}(0)=z\,,\;\; \zeta_{n}(1)=z' \,.
\end{align}

Writing $\zeta^2_{n}$ successively in the form
\begin{align}
\zeta^2_{n}(x)=-\xi^2 x^2 +2\beta x +z^2= a^2-r^2\,,
\end{align}
with
\begin{align}
\xi^2 \equiv 2\ga z z' - z^2 - {z'}^2 \!=\!\om^2\, b^2\,
\sin^2\theta\cos^2\alpha+ b^2 k_T^2=z'^2 \gamma^2 v^2
\sin^2\vartheta \, , \!\quad \beta \equiv \ga z z' - z^2 \, ,
\end{align}
and
\begin{align}
a\equiv \sqrt{\frac{\beta^2}{\xi^2}+z^2} \, , \quad
r\equiv \xi \left(x-\frac{\beta}{\xi^2}\right)\, ,
\end{align}
and using formula f.2.16.12-4 of \cite{Proudn}
\begin{align}
\label{hhh11}
\hat{K}_{\nu-1/2}\!\left(\sqrt{a^2-r^2}\right)=\frac{2^{1/2}}{\pi^{1/2}}
a^{2\nu} \int \limits_0^\infty {\ch}(r y)  \hat{K}_{-\nu}\!
 \left(a \sqrt{y^2+1}\right)dy\,, \quad  \nu>-1 \;{\rm and}\; \,a > 0
\end{align}
for $\mu\!=\!-1/2, \,\nu\!=\!(d-1)/2$ one may rewrite (\ref{cucu1})
in the form
\begin{align}
\label{hhh13}
{J}^{n}(k)= \Lambda_d \frac{2^{1/2} a^{2\nu}}{\pi^{1/2}}
\int\limits_0^\infty dy \, \hat{K}_{-\nu} \! \left(a \sqrt{y^2+1}\right)
\int\limits_0^1 dx\, e^{-i (k b)x} \ch(r y) .
\end{align}
Perform, next, the integration over $x$ and introduce the additional
angle $\phi$, so that the generic $(D-1)$-dimensional unit vector
$\text{\textbf{K}}/|\text{\textbf{K}}|$ (the \emph{normalized}
higher-dimensional emission vector $\bf K$) is decomposed as:
\begin{equation}
\frac{\text{\textbf{K}}}{|\text{\textbf{K}}|}=\frac{\bf{u}}{|\bf{u}|}
\cos\vartheta+ \frac{\mathbf{b}}{|\mathbf{b}|}\sin\vartheta\cos\phi+\bf{m}\sin\vartheta\sin\phi,
\end{equation}
where $\bf{m}$ is a $D-1$ dimensional unit vector orthogonal to
the collision plane (spanned by $\mathbf{u}$ and
$\mathbf{b}$). Then $(k\cdot b)=- \gamma z' v \sin\vartheta
\cos\phi =-\xi \cos\phi$ and $a=\od b \psi /\sin\vartheta$. Substituting this into (\ref{hhh13}) we have
\begin{align}
\label{hhh16}
{J}^{n}(k)\!=\Lambda_d \frac{2^{1/2}
a^{2\nu}}{\pi^{1/2}} \, \frac{1}{\xi} \,\sum_{j=0,1} \!(-1)^{j+1}
e^{-ij (k  b)} \! \!\int\limits_0^\infty \! dy \, \hat{K}_{-\nu}
\! \left(a \sqrt{y^2+1}\right)\frac{y \sh(\coa\delta_j
y)-i\cos\phi \ch(\coa \delta_j y)}{y^2+\cos^2\!\phi} \!\equiv \!
J^{n}_0+J^{n}_1\,,
\end{align}
where $\delta_j=j-\beta/\coa^2, \; j=0,1.$
\footnote{Direct integration of (\ref{cucu1}) for ($\theta\!=\!0$,
\!${\uln}\!=\!0$) gives $\int\limits_{0}^{1}
\hat{K}_{d/2-1}(\sqrt{2\beta x+z^2})\,dx\!=\!
\beta^{-1}[\hat{K}_{d/2}(z)\!-\!\hat{K}_{d/2}(z')]$. The constants
of integration of the terms $j=0,1$ are chosen so that for
$\theta=0$ they satisfy $J^{n}_{z}|_{(\theta=0,{\uln}=0)}=\Lambda_d \beta^{-1}\hat{K}_{d/2}(z);$
$J^{n}_{z'}|_{(\theta=0, {\uln}=0)}=-\Lambda_d
\beta^{-1}\hat{K}_{d/2}(z').$}
The convergence of these integrals is controlled by the
competition of the exponential decay of the Macdonald function and
the exponential growth of the hyperbolic functions. In all cases
the first is faster, but when the difference of the two arguments
is small, the main contribution to the integral over $y$ comes
from large values of $y.$

Since $y^2+1 \geqslant 1$ and $0\leqslant \sin^2\!\phi \leqslant 1$, one can
equivalently write
\begin{align*}
 {J}^{n}_{j}(k)\!=(-1)^{j+1} e^{-ij (k  b)} \Lambda_d
\frac{2^{1/2} a^{\nu}}{\pi^{1/2}} \, \frac{1}{\xi}  \!
\sum_{k=0}^{\infty} \sin^{2k}\!\phi \!\int\limits_0^\infty \! dy \,
\frac{{K}_{\nu} \! \left(a \sqrt{y^2+1}\right)}{(y^2+1)^{\nu/2+1+k}}
\,[y \sh(\coa\delta_j y)-i\cos\phi
\ch(\coa \delta_j y)].
\end{align*}

The $y$-integration for each value of $k$ is performed by successive applications of the identity \cite{GR}
\begin{align}
\label{Mac_reduce}
K_{\nu}(z)=K_{\nu+2}(z)-\frac{2(\nu+1)}{z}K_{\nu+1}(z)
\end{align}
in combination with
\begin{align}
\label{lead0}
a^{\nu}\int\limits_0^\infty \! dy \,\frac{{K}_{\nu+2} \! \left(a \sqrt{y^2+1}\right)}{(y^2+1)^{(\nu+2)/2}}
\left\{
\begin{array}{c}
y \sh(\coa\delta_j y)  \\
\ch(\coa \delta_j  y) \\
\end{array}
 \right\}=\frac{1}{a^2}\left\{
\begin{array}{c}
\coa \delta_j \hat{K}_{\nu+1/2}(z_j) \\
\hat{K}_{\nu+3/2}(z_j)  \\
\end{array}
 \right\}
\end{align}
obtained from (\ref{hhh11}), with argument $z_j=\sqrt{a^2-\xi^2
\delta_j^2}$, i.e. $z_0=z, z_1=z'$.

For example, using (\ref{Mac_reduce}) the $k=0$ term leads to the integrals
\begin{align}
\label{types}
a^{\nu}\int\limits_0^\infty \! dy \, \left[\frac{{K}_{\nu+2} \!
\left(a \sqrt{y^2+1}\right)}{(y^2+1)^{\nu/2+1}}
-\frac{2(\nu+1)}{a} \frac{{K}_{\nu+1} \! \left(a
\sqrt{y^2+1}\right)}{(y^2+1)^{(\nu+3)/2}} \right] \,\left\{
\begin{array}{c}
y \sh(\coa\delta_j y)  \\  \ch(\coa \delta_j y) \\
\end{array}
 \right\}\,.
\end{align}
The first term in the square brackets is given by (\ref{lead0}). The second is computed using again
(\ref{Mac_reduce}), which leads to two new integrals, the first of which is
\begin{align}
\label{lead1}
a^{\nu-1}\int\limits_0^\infty \! dy \,\frac{{K}_{\nu+3}
\! \left(a \sqrt{y^2+1}\right)}{(y^2+1)^{(\nu+3)/2}}  \left\{
\begin{array}{c}
y \sh(\coa\delta_j y)  \\
 \ch(\coa \delta_j y) \\
 \end{array}
 \right\}=\frac{1}{a^4}\left\{
\begin{array}{c}
\coa \delta_j  \hat{K}_{\nu+3/2}(z_j)   \\
\hat{K}_{\nu+5/2}(z_j)    \\
\end{array}
 \right\}\,,
\end{align}
suppressed for $a\gg 1$ compared to (\ref{lead0}). Similarly, the second is further suppressed by
two more powers of $a$. Terms with increasing $k$ are evaluated in the same way and lead to further
suppression by inverse powers of $a^2$.

The end result for $J^n_0(k)$ keeping terms up to $1/a^4$ is then
\begin{align}
\label{J0}
{J}^{n}_0(k)\!=\! \frac{\Lambda_{d}}{a^2 \coa^2}
\left(\! \beta \hat{K}_{d/2}(z)-i
 (kb)\hat{K}_{d/2+1}(z)  -\frac{(d+1)\beta}{a^2}
\hat{K}_{d/2+1}(z) \! + \frac{\beta \sin^2 \!\phi
}{a^2}\hat{K}_{d/2+2}(z) \right)+R_{z}\,.
\end{align}
Notice that $J_0^{n}(k)$ is a series of Macdonalds with argument
$z$. Consequently, it is important mainly in the $z$-region, where
$a=\od b \psi /\sin\vartheta \sim \gamma \gg 1$, a self-consistency check of our approximations.
The coefficients of all
terms in (\ref{J0}) have expansions in powers of $\ga^{-1}$. In
the $z$-region the first term starts with
$\mathcal{O}(\gamma^{-3})$, the second with
$\mathcal{O}(\gamma^{-4})$, the third and fourth terms with order
$\mathcal{O}(\gamma^{-5})$, while the remainder
$R_{z}=\mathcal{O}(\gamma^{-6})$.

Following the same procedure $J^{n}_1(k)$ is written as a series of Macdonald functions with argument
$z'$, namely
\begin{equation}
\label{J1}
{J}^{n}_1(k) \simeq \Lambda_{d} \, \e^{-i (k b)}
\left(\frac{\delta_1}{a^2}\,
\hat{K}_{d/2}(z')-i\frac{\cos\phi}{a^2 \,\xi} \,
\hat{K}_{d/2+1}(z')\right)\,+R_{z'} \,,
 \end{equation}
whose main contribution comes from the region with $z'\sim 1$, i.e.  $\od \sim \ga/b, \vartheta \sim 1$,
in which indeed $a \sim \gamma \gg 1$.

\begin{figure}
\begin{center}
\includegraphics[angle=0,width=9cm]{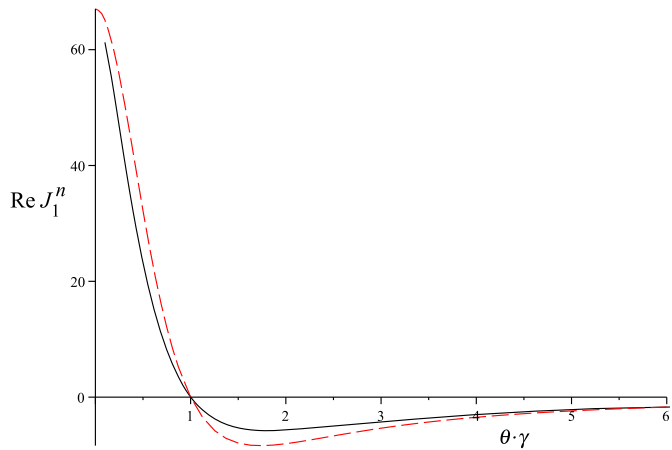}
\caption{The real part of the amplitude $J_1^n$ (solid line)
in $D=4$ dimensions, calculated numerically from the exact formula
(\ref{hhh16}), compared with the approximation given by
(\ref{J1}) (dashed) for $\gamma=1000$ and for $\cos\phi=1$, which gives the largest deviation between the two.
For $\gamma\theta\gtrsim 4$ the deviation is negligible. For $\gamma\theta\lesssim 4$ the exact expression will
be used numerically.
The imaginary part has similar behavior, and furthermore is suppressed compared to the real part by inverse
powers of $\gamma$.}
\label{appr1}
\end{center}
\end{figure}

The condition $a^2\gg 1$ is not satisfied in the region with
$\theta \sim 1/\ga$. However, in that region both the exact
expression and the approximate one have negligible contribution to the
amplitude. Figure \ref{appr1} displays graphically the maximal
difference in the real part of $J^{n}_1(k)$ between the two
expressions.

Incidentally, notice that there is no strong anisotropy in $\phi$ in the $z'-$region: the
 real part of main terms of $J^{n}_1$ (\ref{J1}) is independent on $\phi$, while its
 imaginary part depends only by the overall factor $\cos\phi$. The same picture
 was obtained without any approximations in \cite{GKST-2}, where scalar mediated collisions
 were studied.

Going back to (\ref{Clinton05}) one sees that $\sigma^n(k)$ is the
sum of two sets of Macdonald functions, one with argument $z$ and
the other with argument $z'$. In analogy with $\rho^n$, the
first sum contributes mainly in the $z$-region. Similarly, the
leading contribution of the second set of Macdonalds comes from
the region with $z'=\om b / \ga v  \sim \mathcal{O}(1)$.
In the ultrarelativistic limit this translates into angular and
frequency characteristics of the emitted radiation. For generic
values of the angles, this means $\om \sim \ga/b$ and defines what
we will call the {\it $z'-$region} and, correspondingly, {\it $z'-$type} radiation.
It is unbeamed radiation ($\vartheta \sim 1$) with characteristic frequency $\om \sim \ga/b$.

\vspace{0.5cm}

\textbf{The non-local pieces $\s_0^n(k)$ and $\s_1^n(k)$.} It is
convenient to separate the two kinds of
contributions to the non-local amplitude by writing \begin{equation}\lb{sigman} \sigma^n(k)\equiv \sigma^n_0(k)+\sigma^n_1(k)\,, \end{equation}with the first (second) given by (\ref{Clinton05}) with $J^{n}_0$
($J^{n}_1$) on the right hand side. Thus,
\begin{align}
\label{sigman0}
\s_0^n(k)=
 \lambda\, \e^{i(kb)} \,  \frac{ \ga  v  {z'}^2}{a^2\, \coa^2}
\left(\! \beta \hat{K}_{d/2}(z)-i
 (kb)\hat{K}_{d/2+1}(z)  -\frac{(d+1)\beta}{a^2}
\hat{K}_{d/2+1}(z) \! + \frac{\beta \sin^2 \!\phi
}{a^2}\hat{K}_{d/2+2}(z) \right),
\end{align}
and
\begin{equation}
\label{sigman1}
\s^n_1(k) \simeq \lambda
\frac{\ga v {z'}^2}{a^2\, \xi^2} \left((\xi^2-\beta)\,
\hat{K}_{d/2}(z') + i (kb) \,
\hat{K}_{d/2+1}(z')\right)\,,
 \end{equation}
respectively.

Correspondingly, the total radiation amplitude $j^n(k)$ is written
as a sum of two terms, one function of $z$, and the other function
of $z'$ \begin{equation}\lb{jn} j^n(k) \equiv j^n_z (k)+ j^n_{z'}(k) \, , \;\;
j^n_z(k)=\rho^n(k) + \sigma^n_0(k)\,, \;\;
j^n_{z'}(k)=\sigma^n_1(k) \end{equation}
%%%%%%%%%%%%%%%%%%%%%%%%%%%%%%%%%%%

\subsection{The part $j^n_z(k)$ of the radiation amplitude and destructive interference}

\subsubsection{$j^n_z$ in the frequency range $\omega\gg \ga/b$}

Consider first the regime with $\vartheta\sim 1$. Here $z\sim \ga$ and from (\ref{sigman0}) and (\ref{rhon_final})
one obtains that $j^n_z\sim \exp(-\ga)$ due to the Macdonald functions.

Now take the most interesting case of $\vartheta\sim 1/\ga$, in
which $z\sim 1$. Add $\rho^n(k)$ and $\s^n_0(k)$ given in
(\ref{rhon_final}) and (\ref{sigman0}), respectively, and use the
ultra-relativistic expansions to obtain in leading order:
%\begin{eqnarray}
\begin{align}
\label{jnz}
j_{z}^n(k) \simeq \frac{\lambda\,(d+1) \e^{i(kb)}}{\ga
\psi}\left[ \frac{2 \psi- \gamma^{-2} }{d+2}  \hat{K}_{d/2}(z)
-\frac{\cos^2\alpha}{ \psi^2 \om^2 b^2}\left( \vphantom{\frac{d}{d}} \right.  \right.  & \left(\sin^2\!\theta
+\tg^2\!\alpha \right) \hat K_{d/2+1}(z) \nonumber \\
&- \left. \left. \frac{\sin^2 \theta \sin^2\!\varphi +\tg^2\!\alpha }{d+1}\,{\hat K}_{d/2+2}(z)\right) \right]\,.
\end{align}
%\end{eqnarray}
All terms inside the square brackets are of $\mathcal{O}(\ga^{-2})$.
Given that in the $z$-region $1/\ga \psi = z'/z \sim
\mathcal{O}(\ga)$, the leading contribution to $j^n_z$ above is of
$\mathcal{O}(\ga^{-1})$. Higher order terms have been ignored. The
terms of order $\mathcal{O}(\ga)$ and $\mathcal{O}(1)$, both present
in the ultra-relativistic expansions of $\sigma^n_0(k)$ and
$\rho^n(k)$, have opposite signs and cancel in the sum. This is a
general phenomenon of {\it destructive interference} related to the
gravitational interaction. Thus, the two leading powers in the
ultra-relativistic expansion of the direct $\Phi-$emission amplitude
from the accelerated  charged  particle, cancel the ones coming from
the indirect emission due to the {\it $\Phi-\Phi-h$} interaction. As
a consequence, the $z$-type (beamed and high frequency) part of the
radiation is highly suppressed in the ultra-relativistic limit,
compared to the naive expectation. One can check that destructive
interference is valid also in the case of $\Phi-$radiation in
arbitrary $D-$dimensional Minkowski space-time, which can be
obtained as a limit of the present discussion. It was first observed
for gravitational radiation in $D=4$ \cite{ggm} (using a different
approach) and it was recently generalized to arbitrary dimensions in
\cite{GKST-PLB} . For the system at hand, an alternative proof is
presented in Appendix \ref{DI}, using a different approach also
suitable to the frequency range $\om \gg \ga/b$.

The following comments are in order here: (a) As a check of the
above series of approximations, one may consider the special case
of $\theta=0$, for which $(kb)=-\xi=0$. In this case the exact value of $J^n$ obtained from
(\ref{cucu1}) coincides with the one obtained from the approximate expressions
(\ref{J0}) and (\ref{J1}).  (b) Furthermore, equation
(\ref{jnz}) can be shown to coincide with the corresponding
quantity in the case of non-compactified $D=4+d-$dimensional
Minkowski space. This generalizes to scattering and radiation
processes in the relativistic case, the non-relativistic argument
about the behavior of Newton's potential, i.e. that at distances
$b\ll R$ a point particle generates the $D-$dimensional potential,
while at $b\gg R$ its potential behaves as four-dimensional
\cite{Keha}.

One may equivalently parametrize $j^n_z$ using the angles $\vartheta$ and $\phi$ and write it in the form:
 \begin{align}
 \label{Jo1add}
j_{z}^n(k)= \frac{\lambda\, \e^{i(kb)}}{\ga \psi} \left[\frac{
d+1}{d+2}\,(2 \psi- \gamma^{-2}) \hat{K}_{d/2}(z)- \frac{\sin^2 \!
\vartheta}{ z^2}\left( (d+1)\hat{K}_{d/2+1}(z) -\sin^2\!\phi\,
\hat K_{d/2+2}(z)\right)\right]\,.
\end{align}
Note that in the computation of the emitted energy below both angles will be taken continuous;
a sensible approximation for $N_{\inte}\gg 1$ assumed here.

\subsubsection{$j^n_z$ in the frequency range $\omega\lesssim \ga/b$}

For $\od\ll \ga/b$ and $\vartheta\sim 1/\ga$ using (\ref{rhon_final}) for $\rho^n$ and (\ref{Clinton05}) and (\ref{cucu1})
for $\sigma^n$, one concludes that $|\rho^n|\gg |\sigma^n|$ and, therefore,
\begin{align}
j^n (k) \!\!\left.\vphantom{\frac{a}{a}}\right|_{\,\od \ll
\gamma/b} \simeq \rho^n(k)\simeq -\lambda \left[ \frac{1}{\gamma
\psi} \,\hat K_{d/2}(z) +i\frac{\sin \vartheta \cos\! \phi}{\ga
\psi^2 \od b}\hat{K}_{d/2+1}(z) \right] .
\end{align}
For $\vartheta\sim 1$, on the other hand, $\rho^n$, $\sigma^n_0$ and $\sigma^n_1$ are all of the same order, but
suppressed compared to the previous case. In addition, the contribution of this regime to the emitted energy
will be shown to be further suppressed by the integration measure.

More interesting is the case with $\od\sim \ga/b$.
If $\vartheta\sim 1$, then $z\sim \ga$ and using (\ref{rhon_final}) and (\ref{sigman0}) one concludes that
$j^n_z$ is exponentially suppressed because of the Macdonald functions.
However, for $\vartheta\sim 1/\ga$, one may use (\ref{rhon_final}) and (\ref{hhh16}) to obtain that
$\rho^n\sim \ga$ and $\sigma^n_0\sim \ga$, respectively
\footnote{Using the formulae of Appendix \ref{formulae}
one gets in this kinematical regime $a=z'\ga\psi/\sin\vartheta\sim 1$ and also  $\xi\sim 1$ as well as $\beta\sim 1$.
The integrand in (\ref{hhh16}) is independent of $\ga$. All $\ga$ dependence comes from the overall coefficients
in (\ref{hhh16}) and (\ref{Clinton05}).}.

\subsection{The part $j^n_{z'}(k)$ of the amplitude}

Equation (\ref{sigman1}) can equivalently be written in the form:
\begin{equation}
\label{herakles5_ADD}
j^n_{z'} \simeq - \frac{\lambda}{\ga \psi} \left[\left(\frac{1}{\gamma^2
\psi}-1\right)\hat{K}_{d/2}
 \left(z'\right)+i\frac{\sin \theta \cos \alpha \cos \varphi }{\gamma z' \psi}\hat{K}_{d/2+1}
 \left(z'\right)\right]
 \end{equation}
Furthermore, using the angles $\vartheta$ and $\phi$ it becomes:
\begin{equation}
\label{herakles5_HDM}
j^n_{z'} \simeq - \frac{\lambda}{\ga \psi} \left[\left(\frac{1}{\gamma^2 \psi}-1\right)
\hat{K}_{d/2}
 \left(z'\right)+i\frac{\sin \vartheta \cos \!\phi}{\gamma z' \psi}\hat{K}_{d/2+1}
 \left(z'\right)\right]
 \end{equation}
First, for $\od\gg \ga/b$ one has $z'\gg 1$ and, consequently, $j^n_{z'}$ in (\ref{herakles5_HDM}) is
exponentially suppressed. Next, for ($\od\sim \ga/b$, $\vartheta \sim 1$) $j^n_{z'}$ in (\ref{herakles5_HDM})
is dominated by its real part which is of order $\mathcal{O}(1/\ga)$.
For ($\od\ll \ga/b$, $\vartheta\sim 1$) one obtains $j^n_{z'}\sim \s^n_0 \sim 1/\ga$. As will be shown below,
however, this region contributes negligibly little to the emitted energy. Similarly,
for ($\od\ll \ga/b$, $\vartheta\sim 1/\ga$) on the basis of (\ref{rhon_final}) and (\ref{cucu1}) one concludes that
the amplitude $j^n_{z'}\sim \s^n_0 \ll \rho^n \sim \ga^2$.  Finally, based on numerical study and previous results
in $D=4$ \cite{ggm} one obtains that (\ref{herakles5_HDM}) is valid also in the regime
($\od\sim \ga/b$, $\vartheta\sim 1/\ga$)  and gives $j^n_{z'}\sim \gamma$.

\subsection{Summary}

The behavior of the local and non-local currents in all characteristic
frequency and angular regimes is summarized in the following Table I.

\vspace{0.5cm}

\begin{tabular}{|c|c|c|c|}\hline
  \backslashbox{$\,\vartheta\!$}{$\od$}  &  $\od\ll \gamma/b $ & $
  \od \sim \gamma/b $& $\od \gg
  \gamma/b$ \\ \hline
  $\gamma^{-1}$ & \ths $\begin{array}{l}\text{\small  no destructive interference} \\
   j^n \sim \rho^n \gg \sigma_0^n \sim \sigma_1^n\\   \vphantom{\sigma_0^n \sim \sigma_1^n}  \end{array}$ \ths  &
  \ths $ \begin{array}{l}\text{\small  no destructive interference} \\ j_z^n \sim \rho^n  \sim \sigma^n_0
   \sim \gamma \\ j_{z'}^n =(\ref{herakles5_HDM})\sim \gamma  \end{array}$ \ths &
  \ths $\begin{array}{l} \text{\small  destructive interference} \\j_z^n =(\ref{Jo1add})
  \sim \rho^n/\gamma^2 \sim 1/\gamma\\ j_{z'}^n=(\ref{herakles5_HDM})  \sim \exp(-\gamma)\end{array} $
  \ths  \\[25pt] \hline
  1 &
  \ths$\begin{array}{l}\text{\small  no destructive interference} \\ j^n \sim \rho^n \sim j_z^n\sim j_{z'}^n
  \\ \vphantom{\sigma_0^n \sim \sigma_1^n} \end{array}$ \ths &
   \ths $\begin{array}{l} \text{\small  destructive interference} \\j_z^n =(\ref{rhon_final}+\ref{cucu1})
   \sim \exp(-\gamma)\\ j_{z'}^n=(\ref{herakles5_HDM}) \sim \gamma^{-1}  \end{array}$ \ths  &
   \ths $\begin{array}{l} \text{\small  destructive interference} \\j_z^n =(\ref{rhon_final}+\ref{cucu1})
   \sim \exp(-\gamma) \\ j_{z'}^n=(\ref{herakles5_HDM}) \sim \exp(-\gamma)
    \end{array} \ths$
\\[25pt] \hline
\end{tabular}

\vspace{0.5cm}

\section{The emitted energy - Spectral and angular distribution}

\begin{figure}
\begin{center}
\subfigure[]{\raisebox{1pt}{\includegraphics[width=8.2cm]{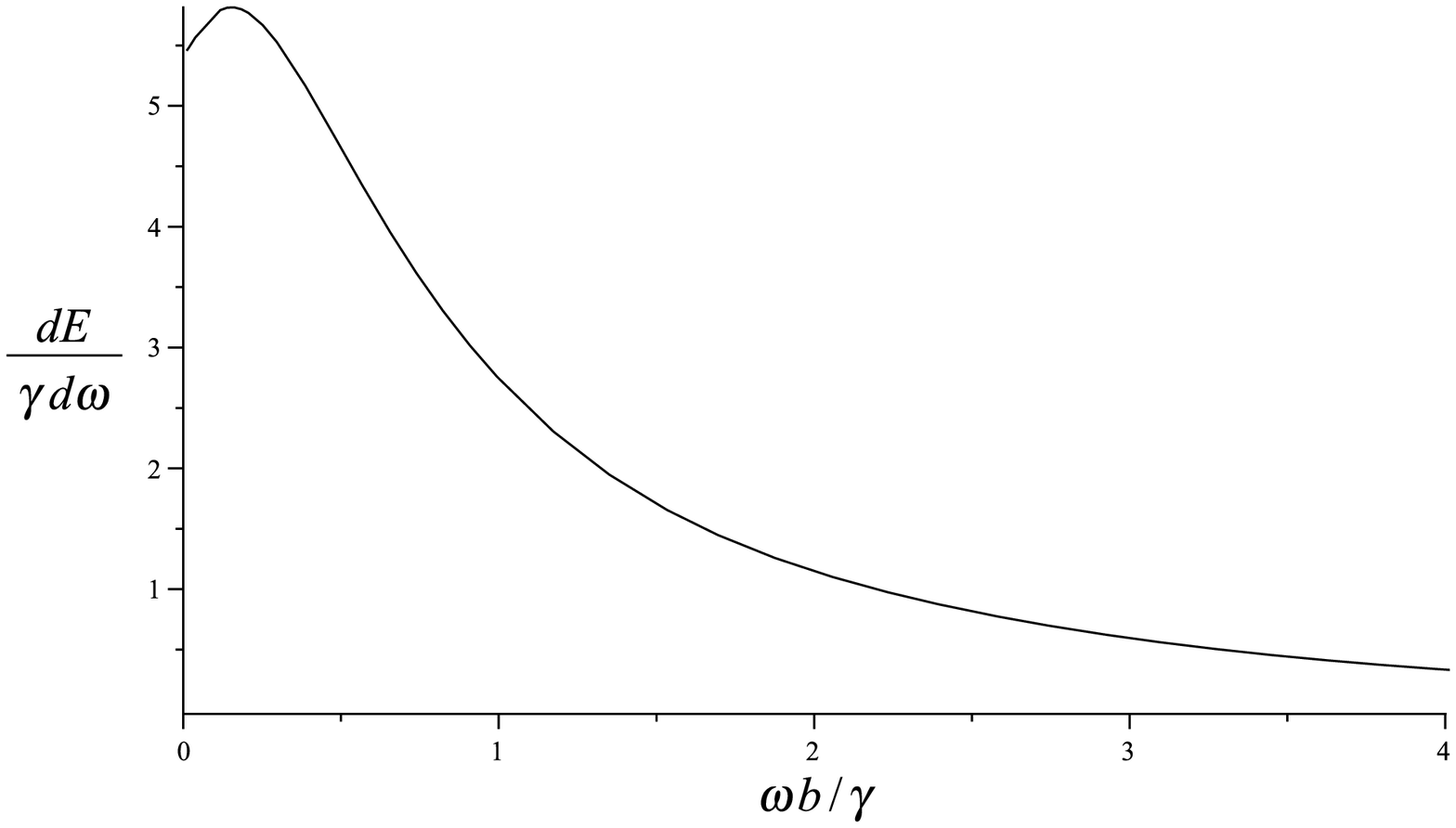}\label{4Ddist_subfig1}}}
\subfigure[]{\includegraphics[width=8.2cm]{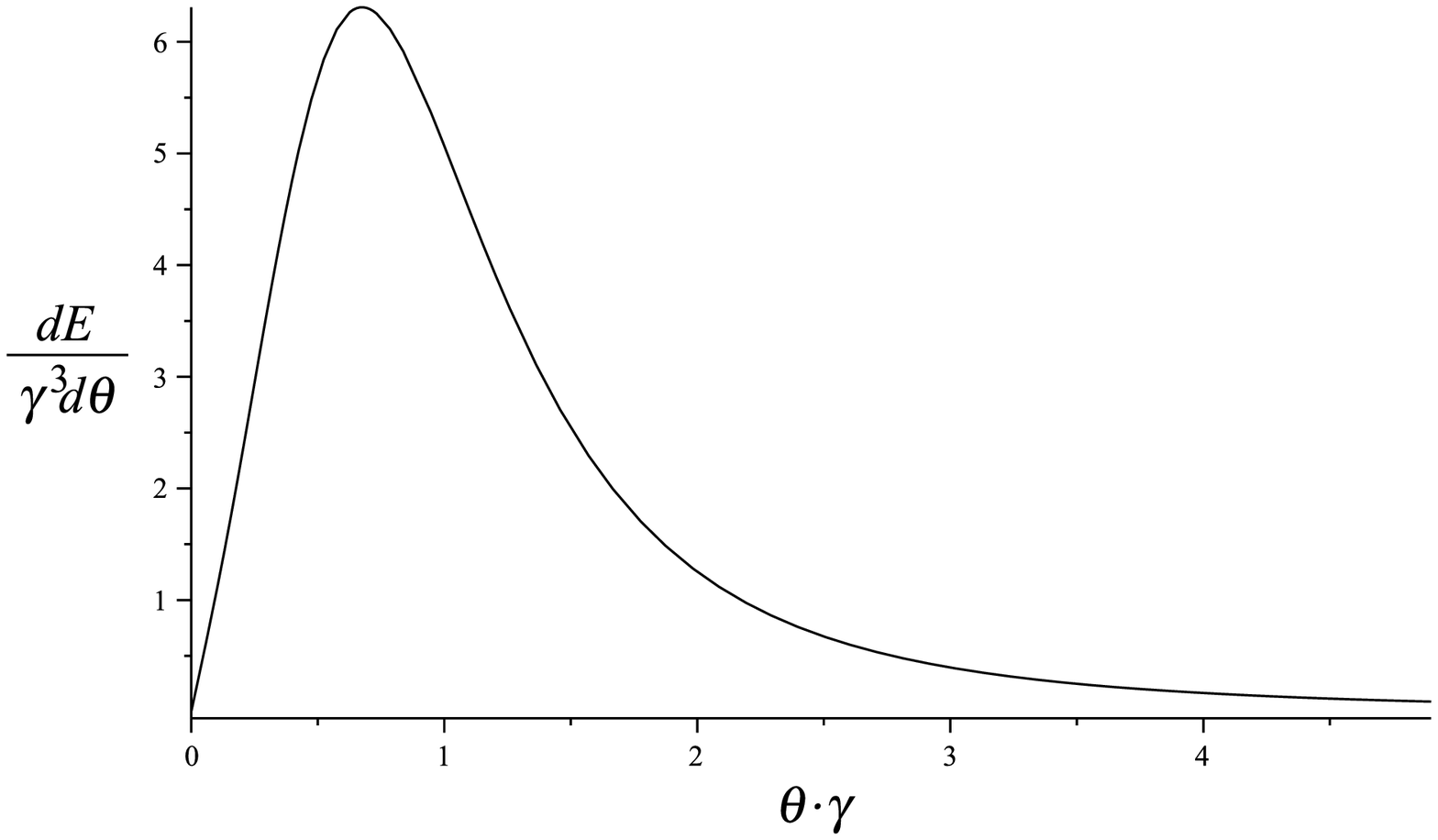}\label{4Ddist_subfig2}}
\end{center}
\caption{Frequency (a) and angular (b) distribution for $d=0$ and
$\gamma=10^5$.}
\label{4Ddist}
\end{figure}

The spectral and angular distribution of the emitted energy is
obtained from (\ref{radiat1}) or (\ref{pafinalmink}). The integrand
is the sum of three pieces proportional to $|j_{z}^n(k)|^2$,
$|j_{z'}^n(k)|^2$ and $\overline{j^n_{z}}  j^n_{z'} + j^n_{z}
\overline{j^n_{z'}}$ (the bar denotes complex conjugation), so the total radiated energy splits into three parts
 \begin{equation}
 dE=dE^{z}+dE^{z'}+dE^{zz'}\,,
 \end{equation}
 which will be called $z-,\,z'-$ and  $zz'-$radiation, respectively.
All terms contain a factor of $\lambda^2$ from the square of the
amplitude, an explicit $1/2(2\pi)^{(d+3)}$ from the Fourier
transform in (\ref{pafinalmink}), and a factor $b^{-(d+3)}$ from the
corresponding power of $\od$ in the integrand. This leads to a
general expression for the total emitted energy of the form
\begin{equation}
E\sim \frac{1}{8(2\pi)^{2d+5}} \frac{\vk^4 {m'}^2 f^2}{b^{3d+3}}\,
\gamma^{\#}\,,
\end{equation}
with an overall coefficient expected to be of
order one and the power of gamma depending on the particular type of
radiation under discussion and which is easily determined as
follows: As argued above and in \cite{ggm} and also shown for
example in Figure \ref{4Ddist} (obtained numerically), the
$\Phi-$radiation is emitted predominantly in well-defined relatively
narrow frequency and angular windows, and with amplitudes shown in
Table I. Thus, it is straightforward to estimate the powers of $\ga$
in the various components of its energy, using (\ref{pafinalmink})

\begin{equation}
E \sim \int d\od \int d\vartheta\, |j|^2\, \od^{d+2} \sin^{d+1}\vartheta
\end{equation}
with $|j|^2$ being $|j^n_z|^2$ or $|j^n_{z'}|^2$ or
$\overline{j^n_{z}}  j^n_{z'} + j^n_{z} \overline{j^n_{z'}}$, and with the range of integration not contributing
extra factors of $\gamma$. For example, the contribution of $j^n_z$, which is dominant in the
regime ($\od \sim \ga^2/b, \vartheta\sim 1/\ga$) has $1/\ga^2$ from $|j^n_z|^2$,
$(1/\ga)^{d+2}$ from the angular integration,  and $(\ga^2)^{d+3}$ from the integration over $\od$, with
the final estimate being $\ga^{d+2}$.

The result of this computation is the content of Table II below~\footnote{Note that in $D=4$ it is found that all three types of radiation are of equal $\mathcal{O}(\ga^3)$.
This seems to disagree with \cite{ggm}, where it is stated that the leading $\mathcal{O}(\ga^3)$ is due to
$z'-$type alone.}.
\\

\begin{tabular}{|c|c|c|c|c|}
  \hline \backslashbox{$\vartheta$}{$\od$} & $\od\ll \gamma/b$  & $ \od \sim \gamma/b$ & $\od \sim \gamma^2/b $& $\od \gg \gamma^2/b$
  \\\hline
  $\gamma^{-1}$ & $\begin{array}{c} \text{\small negligible}\\ \text{\small  (phase space)} \end{array}$ &
  $\begin{array}{l} \hspace{0.1cm} E_d \sim \gamma^3\,, \text{\;\;from $j^n_z$ and $j^n_{z'}$ \hspace{0.0cm}}
  \\ \end{array}$  &
   $\begin{array}{l} \hspace{0.3cm} E_d \sim \gamma^{d+2}, \text{\; from $j_z^n$ \hspace{0.2cm} }\\
   %z'\text{\small - type insig. (exponential fall-off)}
   \end{array}$
   & $\begin{array}{c}  \text{\small  negligible radiation}\\ \end{array}$
  \\[20pt]   \hline
  1 & $\begin{array}{c} \text{\small negligible}\\ \text{\small  (phase space)} \end{array} $ &
   $\begin{array}{l} E_d \sim \gamma^{d+1}, \text{\;\;from $j^n_{z'}$ }\\   \end{array} $&
   $\begin{array}{c}  \text{\small negligible radiation}\\  \end{array}$&
    $\begin{array}{c}  \text{\small  negligible radiation}\\  \end{array} $\\[20pt] \hline
\end{tabular}

\vspace{0.5cm}

We proceed next to the detailed study of the various components of radiation with the frequency and angular
characteristics of the three most important cells of Table II.

\subsection{The $z-$type component of radiation with $\od\sim \gamma^2/b$}

According to Table II, the z-type radiation (due to $|j^n_z|^2$) is always beamed inside $\vartheta\sim 1/\ga$.
Furthermore, for $d\geqslant 2$ it is
dominant with characteristic frequency $\od\sim \ga^2/b$.
The cases $d=0$ and $d=1$ will be treated separately in another subsection.

\begin{figure}
\begin{center}
\subfigure[]{\raisebox{3pt}{\includegraphics[width=8.2cm]{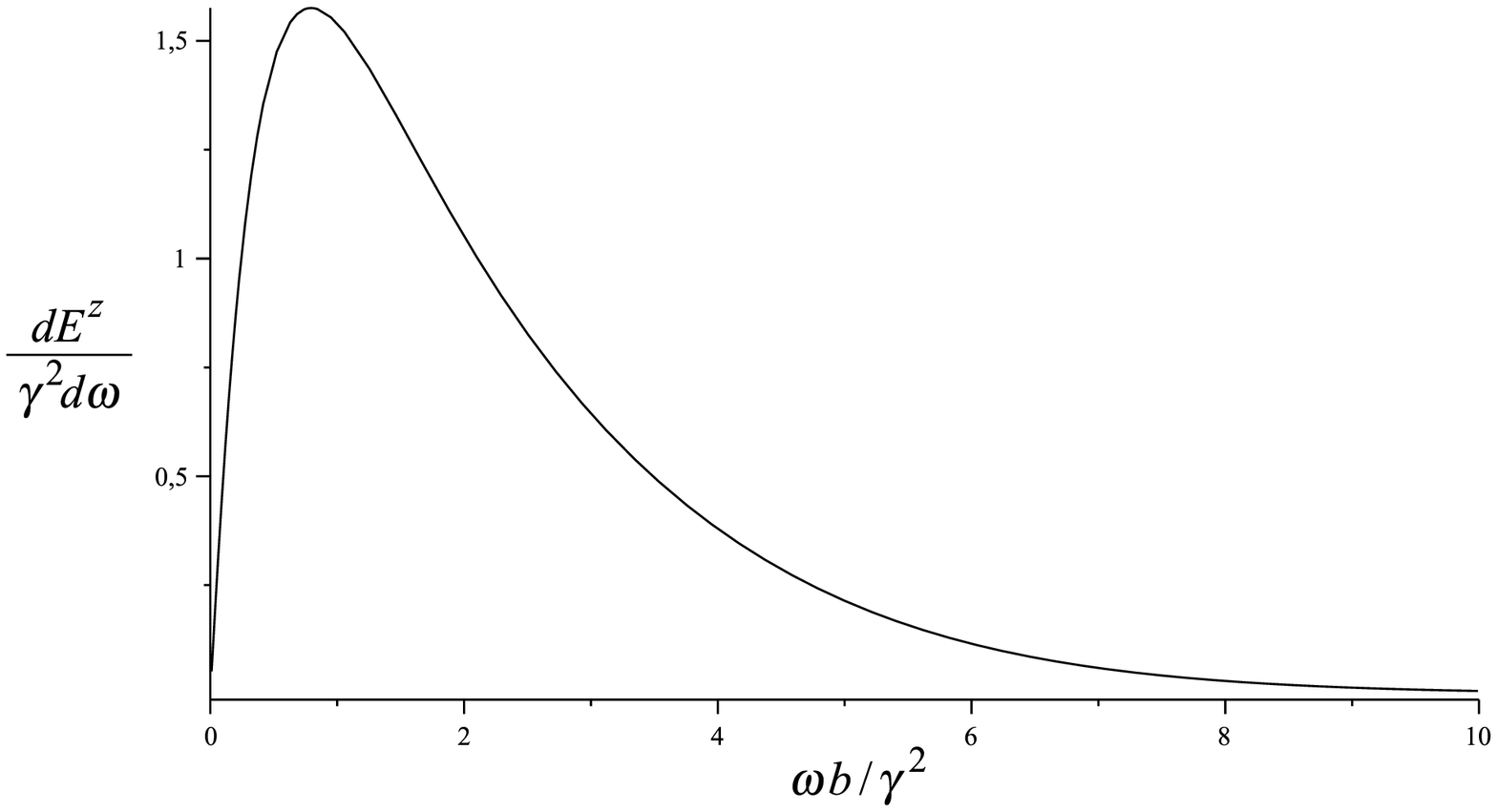}\label{7Ddist_subfig1}}}
\subfigure[]{\includegraphics[width=8.2cm]{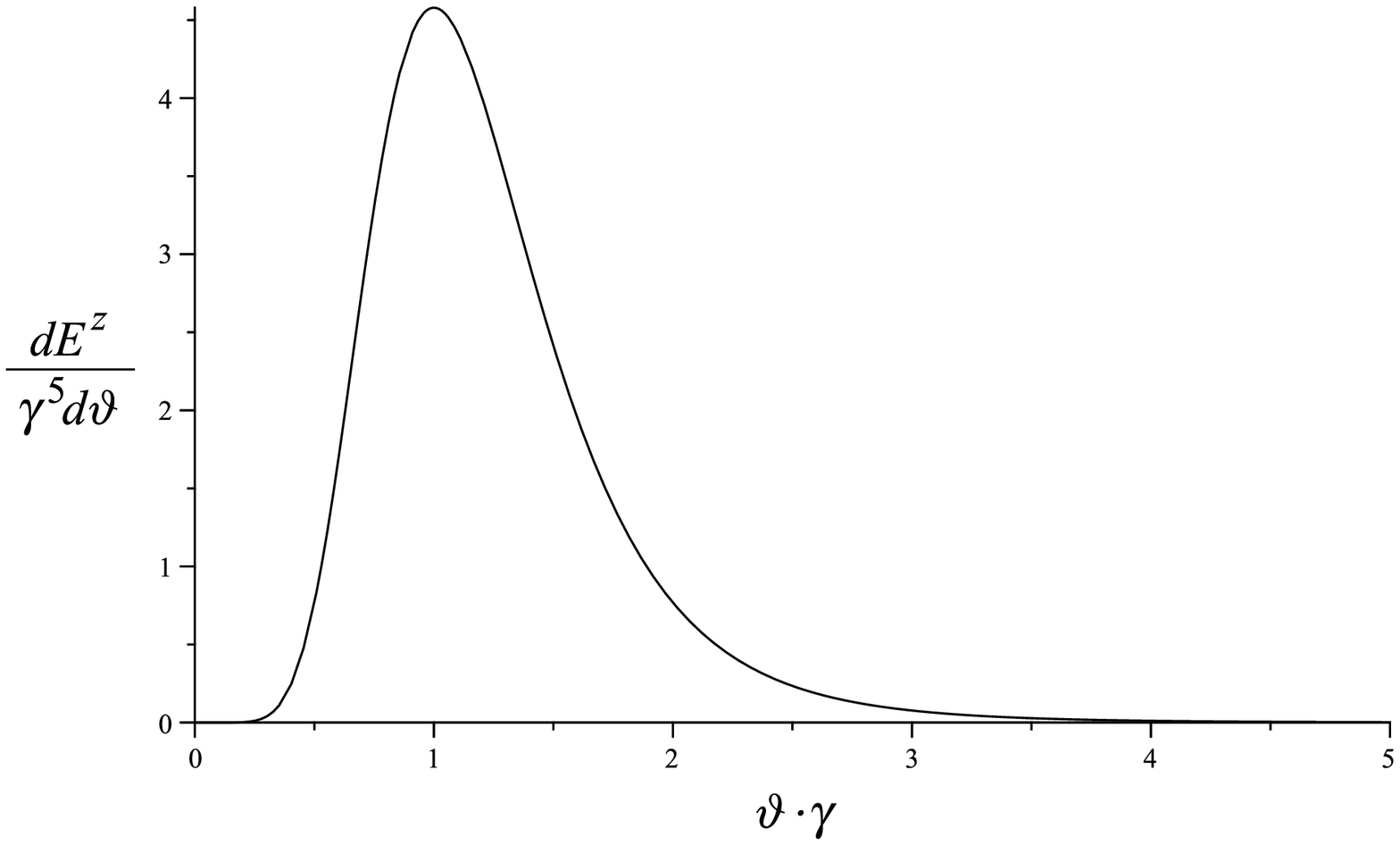}\label{7Ddist_subfig2}}
\end{center}
\caption{Frequency (a) and angular (b) distribution of $z$-radiation
for $d=3$ and $\gamma=10^3$.}\label{7Ddist}
\end{figure}

It is convenient to write the current $j_{z}$ (\ref{Jo1add}) in the form
\begin{align}
\label{Jo012}
&j^n_{z}=\e^{i(kb)}\, \sum_{s=0}^2~^{s\!} j_{z}\\
& ~^{0\!}j_{z}=\frac{\lambda}{\ga}\,  \frac{d+1 }{d+2}\left(2
-\frac{1}{\gamma^{2}\psi } \right)\hat{K}_{d/2}(z)  \\
&~^{1\!}j_{z}=- \lambda \,(d+1)
  \frac{\sin^2 \! \vartheta}{\ga \psi z^2}\hat{K}_{d/2+1}(z) \label{Jo1aaa} \\
&~^{2\!}j_{z}=\lambda \, \frac{\sin^2 \! \vartheta
\sin^2\!\phi}{\ga \psi z^2} \hat K_{d/2+2}(z)
\end{align}
Squaring and substituting into (\ref{pafinalmink}) one obtains
\begin{align}
\label{paracetamol}
\frac{d E^{z}}{d\od d\Omega_{d+2}}= \frac{ \od^{d+2}}{2(2
\pi)^{d+3}}\sum_{a,b=0}^{2} ~^{a\!} j_{z}\!\! ~^{b\!}j_{z}\,.
\end{align}
To integrate over frequencies it is convenient to change variable
from $\od$ to $z$ and define the quantities \begin{equation}\label{Cab}
C_{ab}^{(d)} =\int\limits \hat{K}_{d/2+a}(z)\hat{K}_{d/2+b}(z)
z^{d+2(\delta_{0a}+\delta_{0b}-1)} dz. \end{equation}The integration over
$\vartheta$ is performed using (\ref{jj2q}). Finally, the
integration of $\sin^2\!\phi$ and $\sin^4\!\phi$ over the
remaining angles of $S^{d+1}$ is $\Omega_{d+1}/2$ and
$3\Omega_{d+1}/8$, respectively. The angular and frequency profiles of this component of radiation in
dimensions $d\geqslant 2$, for which it is dominant, were obtained analytically and numerically, respectively, and
have the general form shown for $d=3$ in Figure \ref{7Ddist}.

The end result for the total emitted energy in this component of radiation is
\begin{align}
\label{paracetamol2}
  E^{z}=  \frac{\lambda^2 \Omega_{d+1}}{2(2 \pi\,b)^{d+3}} \,
  \gamma^{d+2}\!\!\sum_{a,b=0}^{2}C_{ab}^{(d)} D_{ab}^{(d)}
\end{align}
where
\begin{align}\label{grebagnecaux}
&D_{00}^{(d)}= \frac{2^{d+1}(d+1)^2 \,\Gamma^2 \!
\left(\frac{d+2}{2}\right)}{ \Gamma( d+4)},& &D_{01}^{(d)}=
-\frac{
 2^{d+3}(d+1)^2\Gamma^2\! \left(\frac{d+4}{2}\right)}{(d+2)\Gamma(d+4)}, &  &D_{02}^{(d)}=-\frac{1}{2(d+1)}D_{01}^{(d)},\nn \\
&D_{11}^{(d)}=   \frac{2^{d+4}(d+1)^2 \Gamma
 \left(\frac{d+6}{2}\right)
 \Gamma \left(\frac{d+4}{2}\right)}{\Gamma(d+5)},  & &D_{12}^{(d)}=- \frac{1}{2(d+1)} D_{11}^{(d)}, & &D_{22}^{(d)}= \frac{3}{8 (d+1)^2} D_{11}^{(d)}\,.
\end{align}

When integrating over $z$ in (\ref{Cab}) one should remember that
the expansion (\ref{Jo1add}) is accurate in the high frequency
domain, around and beyond $z\sim 1$. However, for $d\geqslant 2$
it can be checked both analytically and numerically that the
integral from 0 to 1 of the difference of the exact energy density
based on (\ref{rhon_final}) and (\ref{hhh16}) and the approximate
one based on (\ref{Jo012}) is negligible. Thus, one can
conveniently expand the integration region in (\ref{Cab}) from 0
to $\infty$ and evaluate $C_{ab}^{(d)}$ using (\ref{intfreq}).

Collecting all contributions one obtains for the energy of high frequency $z-$type radiation
\begin{align}
\label{beam_Mink}
E^{z} =C_d \frac{\varkappa_D^4 m'^2 f^2}{b^{3d+3}}\gamma^{d+2}
 \end{align}
with
$C_2=1.42 \times10^{-6}$, $C_3=6.02 \times 10^{-7}$, $C_4=3.45
\times 10^{-7}$, $C_5=2.67 \times 10^{-7}$ and $C_6=2.76 \times
10^{-7}$.

\subsection{ The $z'-$type radiation with $\vartheta\sim 1$}

According to Table II, wide angle radiation ($\vartheta\sim 1$) is mainly $z'$-type (due to $|j^n_{z'}|^2$)
in all dimensions and has characteristic frequency $\od\sim \gamma/b$. Also, for $d\geqslant 3$
radiation with $\od\sim \gamma/b$ is predominantly emitted in wide angles.

\begin{figure}
\centering
 \subfigure[ ]{\raisebox{1pt}{
\includegraphics[ width=8.2cm]{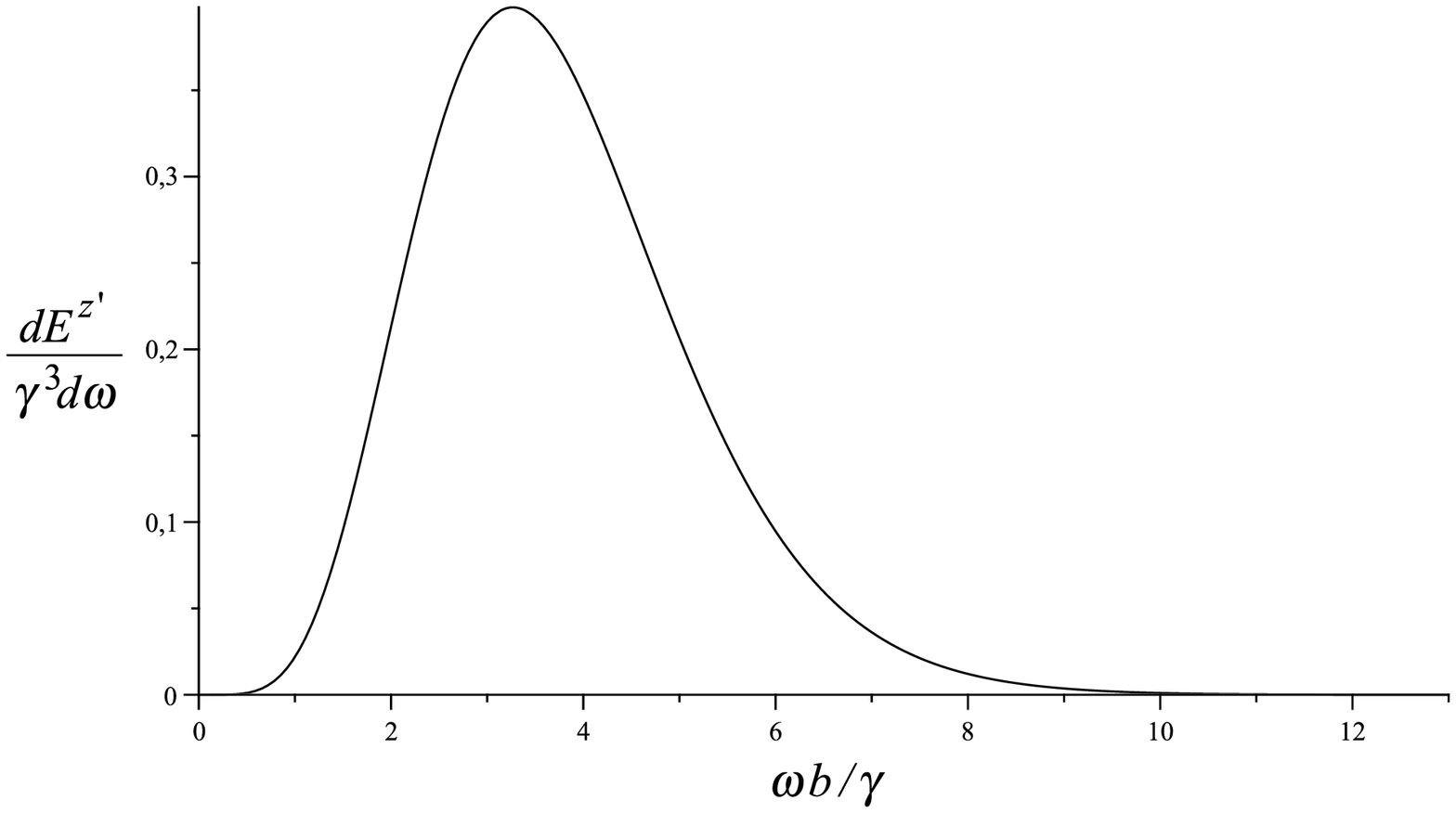}\label{7D_zprime_subfig1}}}
\subfigure[]{ \raisebox{3pt}{
\includegraphics[width=8.2cm]{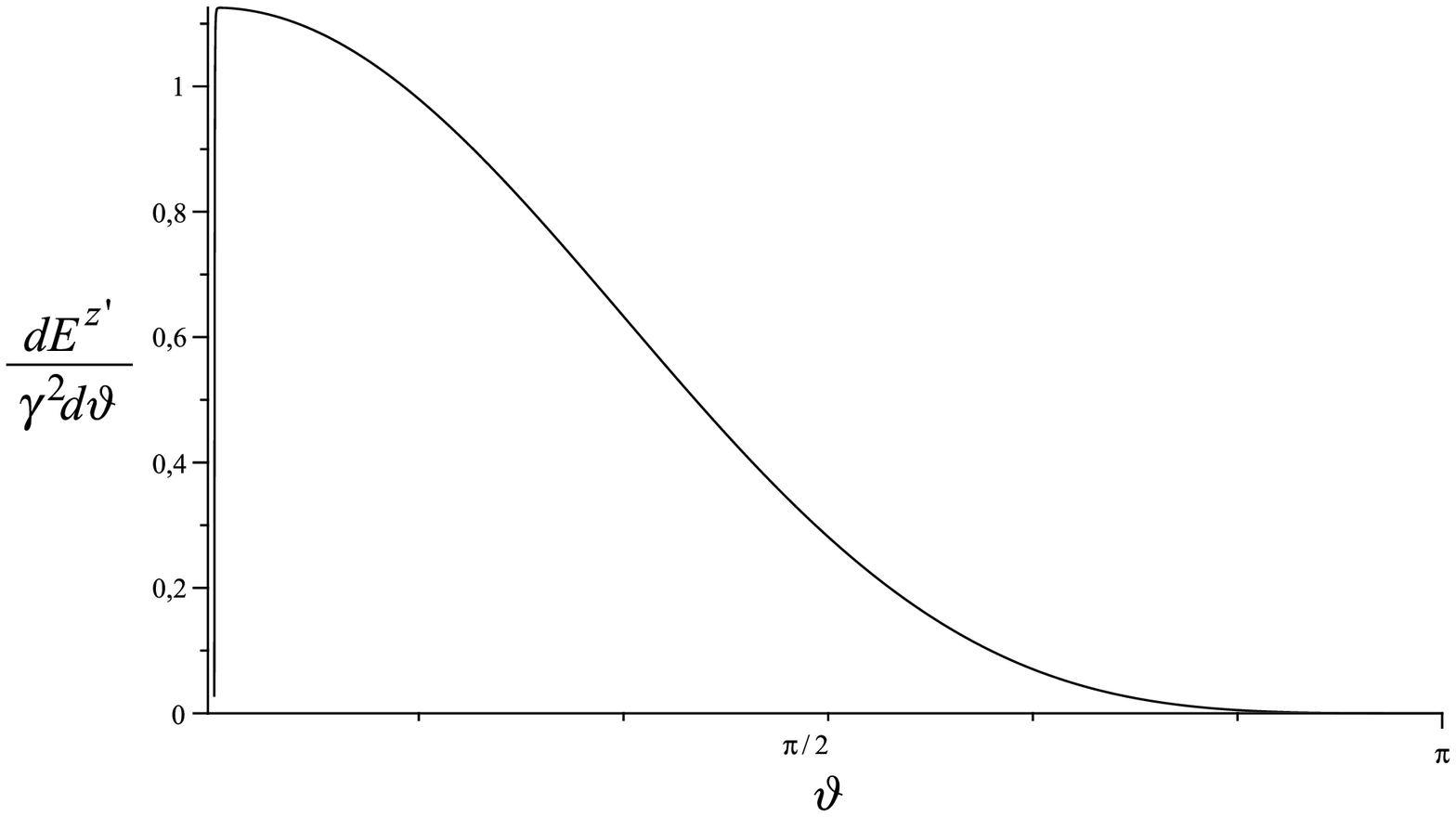}\label{7D_zprime_subfig2}}}
\caption{Frequency (a) and angular (b) distribution of
$z'$-radiation for $d=3$ and $\gamma=10^4$. The angular distribution is actually smooth, but
rises very steeply at this scale for $\vartheta\simeq 0$.}
\label{7D_zprime}
\end{figure}

Squaring
(\ref{herakles5_HDM}), substituting into (\ref{pafinalmink}) and
integrating over $\od$ from 0 to $\infty$ and all angles except $\vartheta$, one gets the angular distribution
\begin{align}
\label{mink_ang_di}
\frac{d E^{z'}}{d\vartheta}=   \frac{\varkappa_D^4 m'^2 f^2
\gamma^{d+1}}{b^{3d+3}} \frac{
  \Gamma\!\left(\frac{3d+3}{2}\right)\Gamma^2\!\!\left( \frac{2d+3}{2}\right)
  \Gamma\!\left(\frac{d+3}{2}\right)}{2^7 \pi^{3d/2+4}
    \Gamma\!\left(\frac{d+2}{2}\right)\Gamma(2d+3)} \frac{\sin^{d+1}\!\vartheta}{ \psi^2}\,.
\end{align}
Formula (\ref{mink_ang_di}) gives the dominant wide angle radiation in all dimensions $d\geqslant 3$.
Figure \ref{7D_zprime} shows the angular and frequency profile of this component
of radiation for $d=3$. To compute the total energy of this type we integrate over $\vartheta$
making use of (\ref{jj3}).

\vspace{0.3cm}

For $d\geqslant 3$ the emitted energy is given by
\begin{align}
\label{en_loss_over7D}
E^{z'}=C'_d\frac{\ka_D^4 m'^2 f^2}{b^{3d+3}}\gamma^{d+1}\,, \qquad
C'_d=\frac{ 2^{d-8}
  \Gamma\!\left(\frac{3d+3}{2}\right)\Gamma^2\!\!\left( \frac{2d+3}{2}\right) \Gamma\!\left(\frac{d+3}{2}\right)
   \Gamma\left(\frac{d-2}{2}\right)}{\pi^{3d/2+4}
  \Gamma(2d+3)  \Gamma(d)}
\end{align}

For $d=2$ one obtains
\begin{align}
\label{en_loss_6D}
 E^{z'}= \frac{105 \varkappa_6^4 {m'}^2 f^2}{2^{16} (2\pi)^7 b^{9}} \,  \gamma^3 \ln \gamma\,.
\end{align}
The cases $d=0, 1$ have to be considered separately since for them
$z$-, $z'$- and $zz'$-types of radiation are comparable and splitting the
amplitude into $j_z$ and $j_{z'}$ is not particularly useful.

\subsection{The cases ${d}$=0, 1}

\begin{figure}
\begin{center}
\subfigure[]{\raisebox{1pt}{\includegraphics[width=8cm]{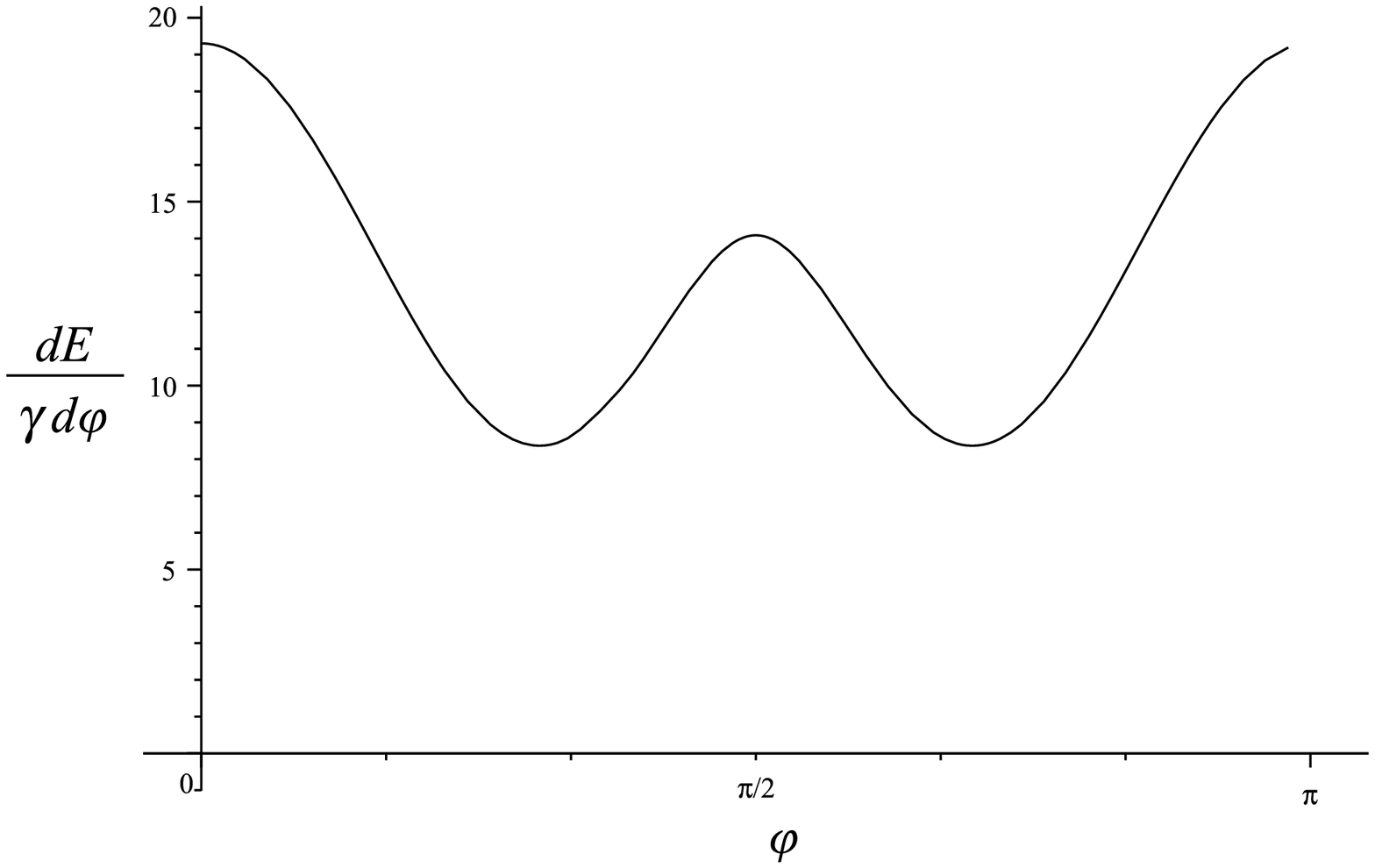}\label{phi_dist_4D_subfig1}}}
\subfigure[]{\includegraphics[width=8cm]{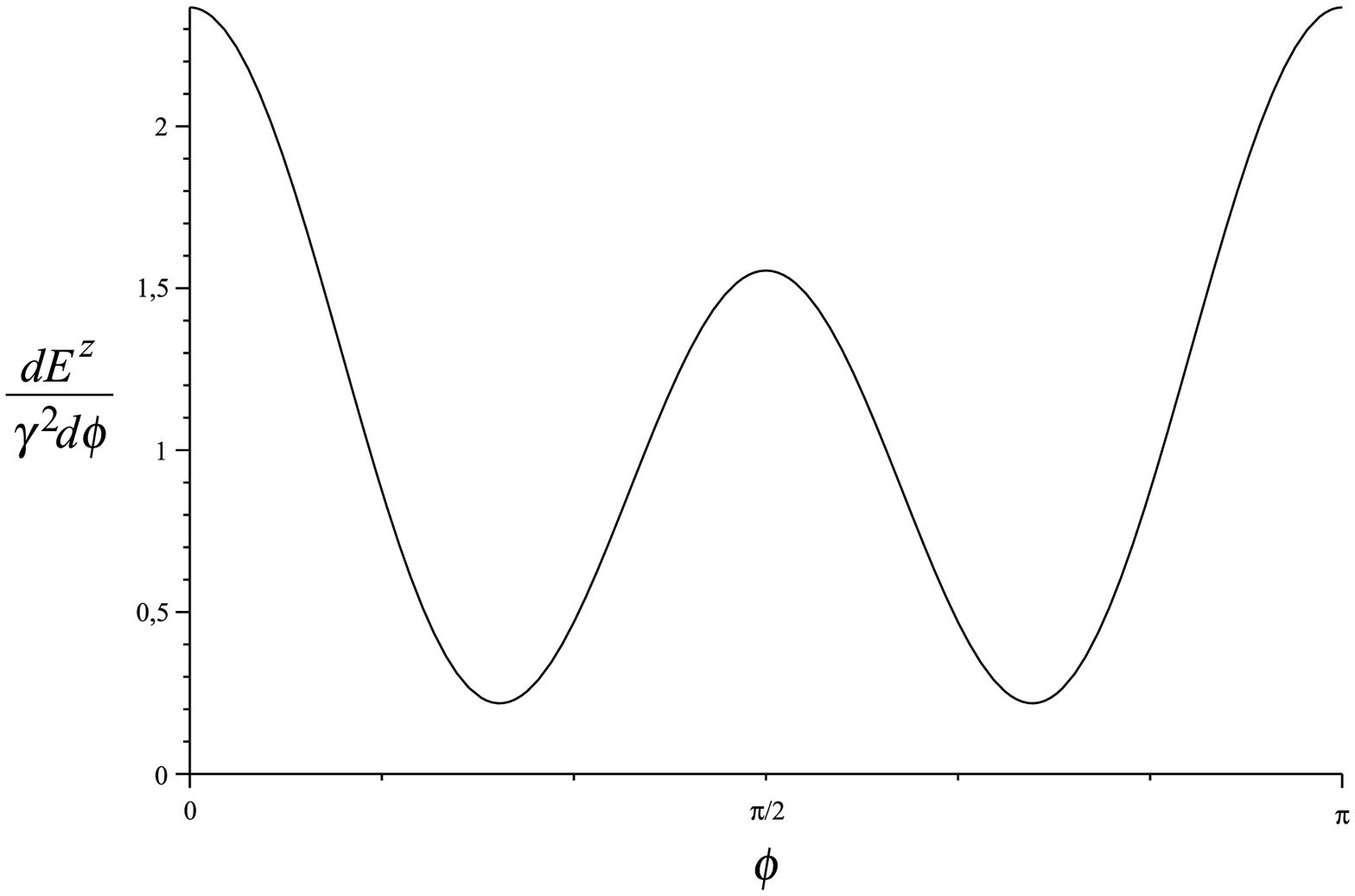}\label{phi_dist_4D_subfig2}}
\end{center}
\caption{(a) The  $\phi-$distribution in 4D for $\gamma=10^4$ and (b) in 6D for $\gamma=10^3$.}
\label{phi_dist_4D}
\end{figure}

According to Table II the emitted energy in 4D is concentrated in
the region $\od \sim \ga/b, \,\theta \sim \gamma^{-1}$.  In this case
the exponent $\e^{i \,x\, \om b \sin \vartheta  \cos \varphi}$ in
the stress amplitude $\sigma(k)$ does not oscillate fast and the
emitted energy may be easily computed numerically. The frequency and
$\vartheta$-distributions in this case are shown for $\gamma=10^5$
in Figure \ref{4Ddist}, while the distribution over $\phi$ (which
coincides with $\varphi$ in 4D) is presented by
Figure \ref{phi_dist_4D}.

The total emitted energy is
\begin{equation}
E_{0}=C_0 \frac{\varkappa_4^4
{m'}^2 f^2}{b^3}\, \gamma^3\,, \qquad C_0 \approx 8.3 \times
10^{-5}\,.
\end{equation}
The frequency distribution is non-zero at $\om=0$ (see Figure \ref{4Ddist}), in
agreement with the analytically derived value
$$\left.\frac{dE_{0}}{d \omega}\right|_{\om=0}=\frac{1}{ 3 \times 2^6 \times \pi^4}\,
\frac{\varkappa_4^4 {m'}^2 f^2}{b^2}\,\ga^2,$$
due mainly to the
imaginary part of the $\rho-$amplitude.

\begin{figure}
\begin{center}
\subfigure[]{\raisebox{2pt}{\includegraphics[width=8.2cm]{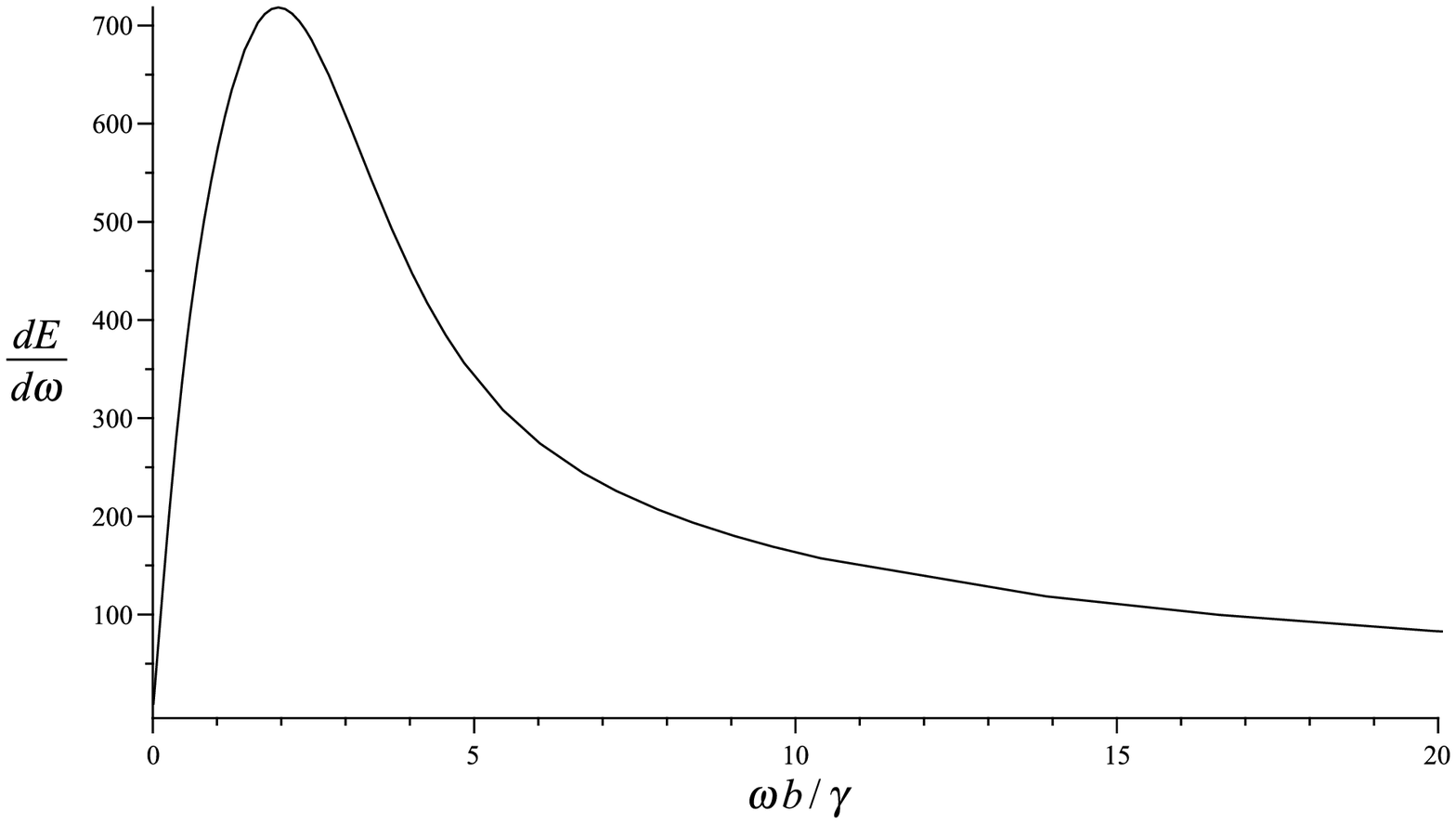}\label{5D_fr_dist_subfig1}}}
\subfigure[\hspace{3pt}  Intermediate region $\od b =20\gamma - 0.1 \gamma^2$: exact distribution (solid black) versus
the approximate $dE/d\omega\sim 1/\omega$ (dashed red)]
{\includegraphics[width=8.2cm]{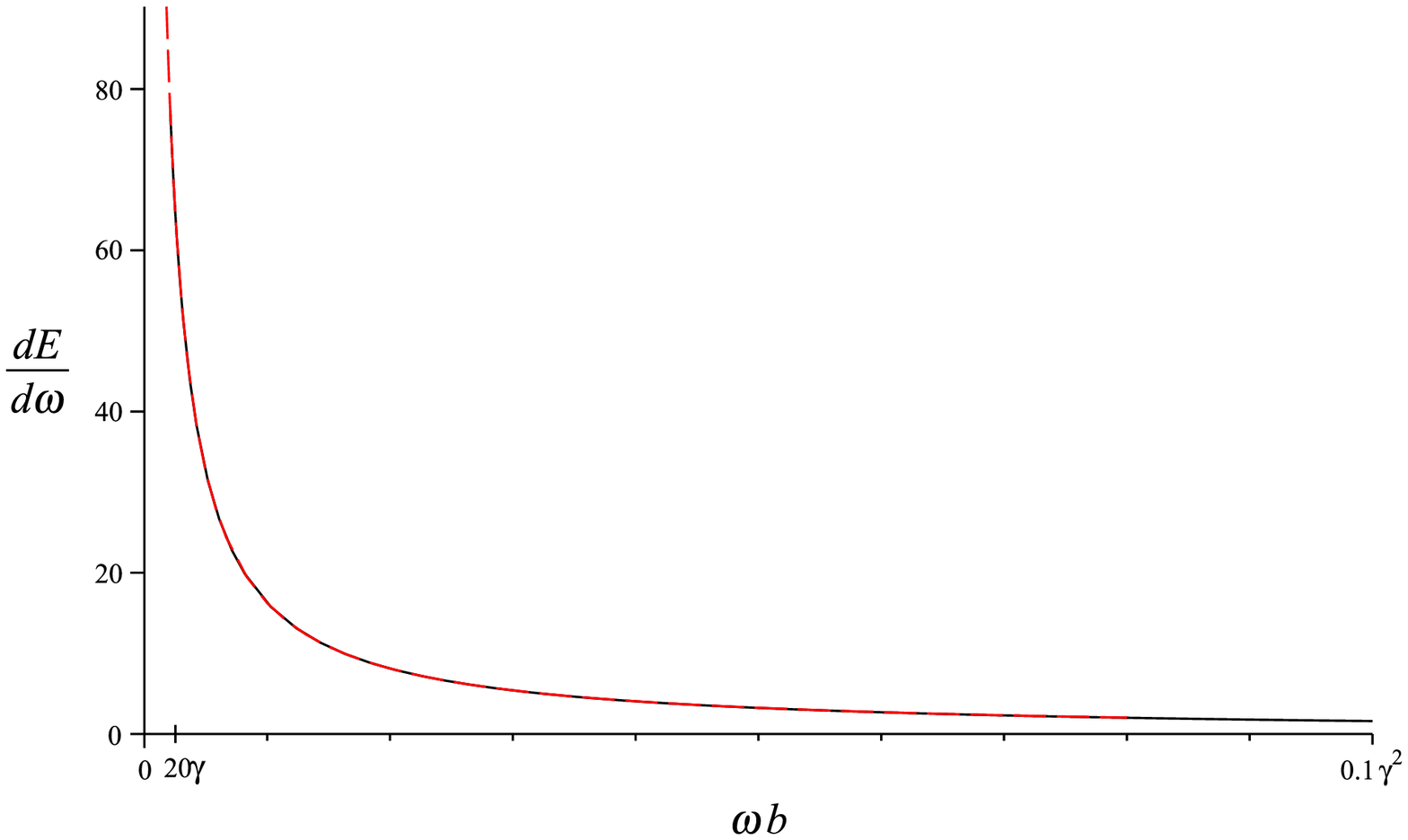}\label{5D_fr_dist_subfig2}}
\end{center}
\caption{Frequency distribution for $d=1$ and $\gamma=10^4$: (a)
for $\od \lesssim \gamma/b$, (b) for $\gamma/b \lesssim \od \lesssim \gamma^2/b$} \label{5D_fr_dist}
\end{figure}

The frequency distribution of the emitted energy $E_1$ in 5D for
$\ga=10^4$ is shown in Figure
\ref{5D_fr_dist}. It is characterized by a long tail beyond the
value $\od\sim \ga/b$, which as can also be argued analytically~\footnote{Notice
from Table I that the total amplitude satisfies $j(\od\sim \ga/b)
\sim \ga^2 j (\od\sim \ga^2/b)$. This gives for $d=1$ the estimate
$|j|^2 \od^{d+2} \sim 1/\od$. }
leads to a behavior $dE_1/d\od \sim 1/\od$ (Figure \ref{5D_fr_dist_subfig2}) all the way to $\od\sim
\ga^2/b$, beyond which it falls-off exponentially. The integral of $dE_1/ d\od$ over the
range ($\ga/b, \ga^2/b$) gives an extra logarithm in the total
emitted energy, which is computed numerically to be
\begin{equation}
E_{1}=C_1
\frac{\varkappa_5^4 m'^2 f^2}{b^6} \gamma^3 \ln \gamma\,,     \qquad
C_1 \approx 1.64 \times 10^{-5}\, .
\end{equation}

\subsection{The estimate of the $zz'-$interference part of radiation}\label{interference}

The purpose of this subsection is to estimate the contribution of the interference
part $(j^n_{z}\, \overline{j^n_{z'}} + c.c.)$. It will be shown that it is subleading for $d\geqslant 2$ and of the same order
as $z-$ and $z'-$ contributions for $d=0, 1$.

The interference term
$E^{zz'}\sim \int (j^n_{z}\, \overline{j^n_{z'}} + c.c.)\, \od^{d+2}\,d\od\, d\Omega_{d+2}$
contains the product of Macdonald functions $\hat K(z) \hat K(z')$.
Thus, its value depends on the overlap of these functions in the
domain $(z\lesssim 1, \, z'\lesssim 1)$, or equivalently $(\od\lesssim \ga/b, \,\vartheta\lesssim 1/\sqrt{\ga})$.
The presence of the factor $\od^{d+2}$ implies that most of the contribution to the integral comes from
the large $\od$ regime with $\od\sim \ga/b$, in which $z'\sim 1$.

For $z \ll 1/\ga$ the integral is suppressed by the volume
factor. Thus, the interesting regime of $z$ is $\gamma^{-1} \lesssim
z\lesssim 1$. In this regime one may estimate the contribution to the interference integral
using
\begin{align}
\label{jzjst}
j_{z} \sim \frac{\lambda \e^{i(kb)}}{2} \,
  \frac{\sin^2 \! \vartheta}{\ga \psi z^2}\hat{K}_{d/2+1}(z)\,, \qquad  j_{z'}  \sim \frac{\lambda}{\gamma
\psi} \hat{K}_{d/2} (z')\,,
\end{align}
since, it can be checked from (\ref{rhon_final}), (\ref{sigman0}) and (\ref{herakles5_HDM}), that they are either
dominant or of the same order as the remaining terms.

One then obtains for the interference part of the energy loss
\begin{align}\label{jzjst2}
\frac{dE^{zz'}}{d\Omega_{d+2} \,d \od}\sim \frac{\la^2 \cos(\od b
\,\sin \vartheta \cos\phi)}{2(2\pi)^{d+3}\gamma^2 \psi^2
z^2}\hat{K}_{d/2+1}(z) \hat{K}_{d/2} (z')\,
\sin^2\! \vartheta\, \od^{d+2}\,.
\end{align}
Integration over all angles except $\vartheta$ gives
\begin{align}
\label{jzjst3}
\frac{dE^{zz'}}{d\vartheta \,d \od}\sim \frac{
\la^2}{2^{d+2}\pi^{(d+3)/2} \Gamma((d+1)/2)\gamma^2 \psi^2
z^2}\,J_0(\od\,b\,\sin\vartheta)\,\hat{K}_{d/2+1}(z)\, \hat{K}_{d/2} (z')\,\,
\od^{d+2}\,\sin^{d+3}\!\vartheta.
\end{align}
The value of the integral over
$\vartheta$ and $\od$ is controlled by $J_0$. For $\vartheta \sim 1/\sqrt{\ga}$, $z$ and $z'$ are both of
$\mathcal{O}(1)$, while the argument of $J_0$
is of $\mathcal{O}(\sqrt{\ga})\gg 1$. Using then the asymptotic expansion of $J_0$ and approximating
the hatted Macdonalds by their values at $z\sim z'\sim 1$, one can estimate as in previous cases the
power of $\ga$ in $E^{zz'}$ to be $\ga^{d/2+3/4}$. This is negligible, compared to the other contributions in all
dimensions.

One is left with the contribution from $\vartheta\sim 1/\ga$, where $\od\,b\,\sin\vartheta\sim 1$ and
$z\sim 1/\ga$. Substituting  $J_0\sim J_0(0)=1$ and $\hat K(z)\simeq \hat K(z=0)$ one estimates
the integral to be of $\mathcal{O}(\ga^3)$ in all dimensions.
This is of the same dominant order as $E^z$ and $E^{z'}$ for $d=0$ and $d=1$ and was included in the total energy
evaluated in the previous subsection.

\section{Summary of results}
Scalar bremsstrahlung radiation during the transplanckian collision of two gravitating massive point
particles in arbitrary dimensions was studied classically in the laboratory frame.
The main goal was to compute the powers of the Lorentz factor $\ga$ and how they depend on the number
of extra dimensions $d$.

We computed both analytically and numerically radiation into truly massless
and massive (for the brane observer) modes. An essential difference
with the previously considered case of scalar bremsstrahlung in flat
space from particles interacting via a scalar field,
is that in the latter the scalar field(s) is linear, while here the bulk scalar interacts non-linearly with
gravity. Within the perturbation theory with respect to
both scalar and gravitational coupling constants it was found that the radiation amplitude
consists of a local and a non-local part. Furthermore, it was shown that in a certain range of angles and
frequencies the leading terms of these two mutually cancel, while the
remaining terms can be presented as the sum of two contributions
($j_z,\,j_{z'}$) which have frequency cut-offs at $\om\leqslant  \ga^2/b$ and
$\om\leqslant \ga /b$, respectively. Their contribution to the total radiated energy $E_d$ depends on the
phase-space in an intricate way, so that the resulting radiation does not have a simple universal expression.
Specifically, it was found that in the absence of extra
dimensions one obtains
 \begin{equation}E_0=C_0\,  m \left(\frac{r_g}{b}\right)^2\left(\frac{r_f}{b}\right)
\gamma^3, \qquad
 C_0 \approx 8.3 \times 10^{-5},
 \end{equation}with the ``basic'' relativistic enhancement factor $\ga^3$. For one
extra dimension one has
 \begin{equation} E_1 =C_1 \, m
\left(\frac{r_g}{b}\right)^4\left(\frac{r_f}{b}\right)^2 \gamma^3
\ln \gamma,
 \qquad C_1 \approx 1.64 \times 10^{-5}\,,
 \end{equation}with almost the same (up to the logarithm) enhancement factor. For $d\geqslant 2$ one finds
 \begin{equation} E_d =C_d \, m
\left(\frac{r_g}{b}\right)^{2(d+1)}\left(\frac{r_f}{b}\right)^{d+1}
\gamma^{d+2} \,,
 \end{equation}
 with $C_d$ are computed from (\ref{paracetamol2}), (\ref{Cab}) and (\ref{grebagnecaux}) and given above for
 $d=2,\ldots, 6$.  So the expected enhancement factor $\gamma^{d+2}$ is regained,
with each new dimension adding one power of $\ga$ to the radiation loss.

Another feature of interest is the spectral-angular distribution of
radiation. It was shown that in the usual gravity theory without
extra dimensions the partial cancelation of local and non-local
amplitudes in the case of gravitational interaction can be
attributed to the fact that in terms of a curved space picture the
world lines of a massive ultrarelativistic radiating charge and the
null geodesic of the emitted radiation stay close to each other, so that the
formation length of the radiation emitted predominantly in the forward direction is $\ga$
times stronger than in flat space. In perturbation theory
on a flat background this corresponds to cancelation of contributions
at high frequencies. In the presence of extra dimensions the emitted
radiation is predominantly massive from the brane observer point of
view, so the trajectories do not stay close together. The resulting
spectral distribution then has substantial remainder at high
frequencies up to $\ga^2/b$, as illustrated in Figs \ref{7Ddist_subfig1}, \ref{5D_fr_dist_subfig2}.

The problem considered in this paper is a simplified intermediate step towards the fully gravitational
counterpart, where one is interested in gravitational bremsstrahlung in particle collisions
interacting gravitationally \cite{GKST-PLB}. The non-local part of the amplitude in this case is
due to the three-graviton vertex. The details of this problem is the subject of a forthcoming publication.

\section*{Acknowledgements}
Work supported in part by the EU grant
FP7-REGPOT-2008-1-CreteHEPCosmo-228644 and 11-02-01371-a of RFBR. DG
and PS are grateful to the Department of Physics of the University
of Crete for its hospitality in various stages of this work. TNT
would like to thank the Theory Group of CERN, where part of this
work was done, for its hospitality. Finally, we are grateful to G. Kofinas for useful discussions.

\appendix

\section{Notation}
\subsection{KK mode decomposition and Fourier transformation - Notation and conventions}
\label{KKnotation}

The Fourier decomposition of the bulk fields $ h_{MN}(x^\mu, y^i), \,
h'_{MN}(x^\mu, y^i),\,\Phi(x^\mu, y^i)$ all with
periodic conditions e.g. $h_{MN}(x,y^k+2\pi R)=h_{MN}(x,y^k)$
is of the form
 \begin{equation}
   h_{MN}(x,y)=\frac1{V}\sum_{n_1=-\infty}^{+\infty} \dots
\sum_{n_d=-\infty}^{+\infty}  {h^n_{MN}(x)} \exp\left( i\frac{n_i
y^i}{R}\right) \equiv  \frac1{V}\sum_{n} h^{n}_{MN}(x) \e^{i n_k
y^k/R  }.
\end{equation}
Using the representation of the delta-function
\begin{equation}
\label{deldisc}
\frac{1}{V}\sum_{n}
\e^{i n_k (y^k-y'^k)/R}  = \delta^d(\y-\y'), \qquad
\frac{1}{V} \int\limits_V\e^{in_ky^k}dy^d=\prod_{k=1}^d \delta_{n_k,\,0}\,,
\end{equation}
 where $V =(2\pi R)^d$ is the volume of the torus, one obtains the
inverse transformation:
 \begin{equation}
 h^n_{MN}(x)=  \int_V h_{MN}(x,y)
  \e^{-i n_k y^k/R} d^dy\,.
 \end{equation}
Four-dimensional fields $h^{n}_{MN}(x)$ are then expanded
as
 \begin{equation}
 h^{n}_{MN}(x)=\frac1{(2\pi)^4} \int \e^{-i(px)} h^{n}_{MN}(p)\, d^4p,
 \end{equation}
 where $(px) \!=\!px\! =\! q_\mu x^\mu$ is four-dimensional scalar product, and
the final decomposition reads
 \begin{equation}
  h_{MN}(x,y)=\frac1{(2\pi)^4}\frac1{ {V }}\sum_{n }\int
h^{n}_{MN}(p) \e^{-i(px)+i n_k y^k/R  }d^4 p\,,
 \end{equation}
while the inverse transformation is
 \begin{equation}
  h^{n}_{MN}(p)= \int\limits_V d^dy \int\limits_{R^4}  h_{MN}(x,y)
\e^{i(px)-i n_k y^k/R } d^4 x.
 \end{equation}
Occasionally we will also use another notation for the discrete
transversal momenta: $p_T^i=n^i/R$, i.e.
 \begin{equation}
 h^{n}_{MN}(p)= \int   h_{MN}(x,y) \e^{i(px)-i {p_T} y  } d^D x.
 \end{equation}
with ${p_T} \, y=p_T^i y^i$. From the four-dimensional point of
view the zero mode $h^0_{MN}(x)$ ($n=0$ means all $n^i=0$) is
massless, while the $n\neq 0$ modes are massive. Indeed, in the
absence of the source term Eq. (\ref{Einm}) reduces to
 \begin{equation}
  (\square+p_T^2)\,h^{n}_{MN}(x)=0,\quad p_T^2=\frac1{R^2}\sum_{i=1}^{d}
 (n^i)^2\,,
 \end{equation}
where $\square=\pa_\mu\pa^\mu$ is the four-dimensional D'Alembert operator, while
the momenta transverse to the brane give rise to the mass term. In
the standard scheme \cite{GRW} one suitably combines polarization
modes to get true massive gravitons with five spin states for each
mass. For our purposes it will   be easier to sum over modes using the original decomposition.

When the level spacing is small (e.g. when $R\gg b$), one
can pass from summation over $n$ to integration over $p_T$ using
\begin{equation}
\lb{sumint} \frac{1}{V} \sum_n=\frac{1}{(2\pi)^d}\int d^d {p_T} .
\end{equation}
Here it is implicitly assumed that both the sum and the integral  converge. As pointed out in the text,
and in contrast to quantum Born amplitudes, this is guaranteed in the framework of the classical
perturbation approach presented here \cite{GKST-EL}.

It is worth noting, that upon integration over modes in the case of
small level spacing, one obtains the results expected in the uncompactified
theory in $D=4+d-$dimensional Minkowski space.

In a similar fashion, expansion of $h'_{MN}(x,y)$ leads to the set of four-dimensional
modes ${h'}^n_{MN}(x)$, and an expansion of the bulk scalar
$\Phi(x,y)$ to the set $\Phi^n(x)$. These four-dimensional fields
are further Fourier transformed to ${h'}^n_{MN}(p)$ and $\Phi^n(p)$,
respectively.

\subsection{Integration over angles and frequencies}
\label{integrals}

In the main text the following integrals over the radiation angle
$\theta$ were encountered
\begin{align}\label{jj0}
V_{m}^n=\int\limits_0^{\pi}\frac{\sin^n
\theta}{\psi^m}d\theta,\qquad \psi=1-v \cos{\theta}
\end{align}
with integers $m,n.$

 For $2m>n+1$ one finds to leading order \cite{GKST-2}
\begin{align}
\label{jj2q}
V_{m}^{n}= & \frac{2^{m-1}}{\Gamma(m)}    \Gamma
 \left(\frac{n+1}{2}\right)
 \Gamma \left(m-\frac{n+1}{2}\right)
\gamma^{2m -n-1} .
\end{align}

For $n>2m-1 $ one obtains
\begin{align}
\label{jj3}
V_{m}^{n}= \frac{2^{n-m} \Gamma\left(\frac{n+1}{2}\right)
\Gamma\left(\frac{n+1}{2}-m\right)} { \Gamma(n -m+1
)}.
\end{align}
In the case $2m=n+1$ the leading contribution to the integral is proportional to $\ln\ga$.
For example, one case needed in the text was $V^3_2\simeq 4\ln\ga$.

Calculation of the integrals over the frequency or over the impact
parameter involving two Macdonald functions of the same argument
is performed using the formula  \cite{Proudn}:
\begin{align}
\label{intfreq}
 \int\limits_0^{\infty}K_{\mu}(cz)K_{\nu}(cz)z^{\alpha-1}dz=
\frac{2^{\alpha-3}\Gamma \left(\frac{\alpha+\mu+\nu}{2} \right)
\Gamma \left(\frac{\alpha+\mu-\nu}{2} \right)\Gamma
\left(\frac{\alpha-\mu+\nu}{2} \right)\Gamma
\left(\frac{\alpha-\mu-\nu}{2} \right)}{c^{\alpha}
\Gamma(\alpha)}.
\end{align}

\subsection{Useful kinematical formulae}
\label{formulae}

The angles in the formulae below are defined in Figure \ref{branepic}.
\begin{align}
&u^\mu\!\equiv \ga(1, 0, 0, v) \,, \;\; u'\equiv (1, 0, 0, 0)\,, \;\;
\psi\equiv 1-v\cos\theta\cos\alpha=1-v\cos\vartheta \,, \;\;  \nonumber \\
&z'\!=\frac{(ku')b}{\ga v}\!=\!\frac{\om b}{\ga v  }\,, \;\;
z\!=\!\frac{(ku)b}{\ga v}\!=\!\frac{\om b}{v }\, \psi=z' \ga \psi\,,
\;\;
 \nonumber \\
&2\ga z z'-z^2-{z'}^2=\om^2 b^2  {\sin^2\vartheta} \,, \;\;
\xi^2\equiv 2\ga z z'-z^2-{z'}^2
=\om^2 b^2 \sin^2\theta\cos^2\alpha + b^2 k_T^2 = (\ga v z' \sin\vartheta)^2 \nonumber \\
&-(kb)=\xi \cos\phi =
\gamma z' v \sin \vartheta \cos\phi =\gamma z' v\cos \alpha
\sin \theta \cos \varphi =\omega b \sin \theta \cos \varphi        \nonumber \\
&\beta\equiv \ga z z' - z^2 = \frac{\om^2 b^2 \cos\vartheta
(1-v\cos\vartheta)}{v  } =\ga^2 {z'}^2 \psi (1-\psi)
\end{align}

\section{Destructive interference for $\gamma/b\lesssim \om \lesssim \gamma^2/b$
\label{DI}}

An alternative proof of the destructive interference effect of the radiation amplitude
in the $z-$region but with $\vartheta < 1/\gamma$ in higher dimensional Minkowski
space will be presented here. In the main text we followed an approximation allowing to cover the full angular range.
Here destructive interference  in the restricted angular range will be demonstrated rigorously.

Start with (\ref{cucu1}), change variable $x$ to $\zeta$ given by
\begin{equation}
\zeta \,d\zeta = f(x)\, dx\, , \quad f(x)=(z^2+{z'}^2-2\ga z z')\, x+\ga z z'-z^2 \,,
 \end{equation}
and integrate by parts twice using
\begin{equation}
\zeta{\hat K}_\nu(\zeta)=-{\hat K}'_{\nu+1}(\zeta)\,.
\end{equation}
The first integration gives:
\begin{equation}
\int\limits_0^1 dx \,e^{- i x (k  b)} \hat{K}_{d/2 -1}\left[\zeta
(x)\right]
=-\frac{\e^{- i x (k  b)}}{f(x)} \,\hat{K}_{d/2 } \left[\zeta (x)\right]\, \Big|_{x=0}^{x=1}
+ \int\limits_0^1 dx \,  \hat{K}_{d/2}(\zeta) \, \pa_x\left(\frac{\e^{- i x (k b)}}{f(x)}\right).
\end{equation}
A second integration by parts leads  to
\begin{align}
\label{bypart}
\s(k) &=\la_d \frac{\gamma\, v\,{z'}^2}{ b^d }\Bigg[\fr{e^{ i  (k  b)}}{\ga z z'-z^2}\!\left(\!
 \hat{K}_{d/2}(z)\! - \! i \,q_0 \frac{\hat{K}_{d/2+1}(z)}{\ga z z' - z^2}
\!\right) \! - \! \fr{1}{{z'}^2-\ga z z'}
\!\left(\!\hat{K}_{d/2}(z') \! - \! i \,q_1\frac{\hat{K}_{d/2+1}(z')}{{z'}^2-\ga z z'}
\!\right)\! +\! R\Bigg] ,
 \end{align}
where
\begin{equation}\lb{betas}
q_0=(kb)-i\fr{{z}^2+{z'}^2-2\ga zz'}{\ga zz'-z^2},\quad
q_1=(kb)-i\fr{{z}^2+{z'}^2-2\ga zz'}{{z'}^2-\ga zz'},
\end{equation}
and
\begin{equation}\lb{cde}
R=\int\limits_0^1 dx \, \hat{K}_{d/2 +1}(\zeta(x))  \left[ \left(\frac{\e^{- i x (kb)}}{f(x)}\right)^\prime \frac{1}{f(x)}\right]^\prime.
\end{equation}
Continuing integration by parts further, one obtains an expansion
in terms of $q_0 \beta^{-1}$ and $q_1
(\beta-\coa^2)^{-1}$. As
we discussed before, in the $z-$region of interest here $\psi\sim 1/\ga^2,\; z\sim 1
,\; z'\sim \ga,$ so that $\;
\coa^2\sim \beta\sim\ga^2 \sim  (\beta-\coa^2), \; q_0\sim q_1\sim\ga$ and therefore the expansion
parameters are: $q_0 \beta^{-1}\sim \ga^{-1}\ll
1,\;q_1 (\beta-\coa^2)^{-1}\sim \ga^{-1} \ll 1$. With this accuracy one can set $q_0=q_1=(kb),\; \beta=\ga zz'$
and write:
\begin{align}
\label{nonlocfin}
 \s(k) \simeq \frac{\la_b}{b^d }\Bigg[ e^{ i  (k  b)}\!\left(\!
\frac{z'}{z} \hat{K}_{d/2}(z)- i \frac{(kb)}{\ga z^2}\,\hat{K}_{d/2+1}(z)
\!\right)\!  + \!\left(\!\frac{z'}{z-z'/\gamma}
 \hat{K}_{d/2}(z')+ i \,\fr{(kb)}{\ga z^2}\, \hat{K}_{d/2+1}(z')
\!\right) \Bigg] .
 \end{align}
The first parenthesis in $\sigma$ cancels for $v=1$ the leading terms of $\rho$
(\ref{rhon_final}), and the total amplitude $j(k)=\rho(k)+\s(k)$ contains
only the second parenthesis in (\ref{nonlocfin}) plus the subleading terms mentioned above.
Thus, the series obtained by integration by parts, converges inside $z-$cone $\vartheta<\arcsin \gamma^{-1}$
and establishes the effect of {\it destructive interference}.

\end{document}